\documentclass[11pt,paper=A4]{article}
\usepackage[DIV=12]{typearea}
\usepackage[T1]{fontenc}
\usepackage[utf8]{inputenc}
\usepackage[american]{babel}
\usepackage{mathtools} 
    \mathtoolsset{centercolon} 
    \allowdisplaybreaks[3]

\usepackage{microtype}
\usepackage{amsthm}
\usepackage{amssymb}

\usepackage{tikz} 
\usetikzlibrary{calc,intersections}
\tikzset{%
    myLine/.style={line width=#1*0.89pt},
    myLine/.default=1}

\usepackage{booktabs}
\usepackage[
    hidelinks,
    bookmarksnumbered,
    bookmarksopen,
    bookmarksopenlevel=2,
    bookmarksdepth=3,
    pdfusetitle=true,
]{hyperref}
\newcommand{\statement}[1]{\paragraph{#1}\pdfbookmark[0]{#1}{#1}} 
\usepackage{orcidlink}

\usepackage{caption}
\usepackage{floatrow}
\usepackage[llbracket,rrbracket]{stmaryrd}

\usepackage{stackengine} 
\stackMath 

\makeatletter
\newcommand{\arrow@text}[1]{\scriptstyle\mathrm{#1}}
\newcommand{\labeledRightArrow}[1]{ %
  \mathrel{ 
    \stackon[-3.5pt]{ 
      \xrightarrow{\hphantom{\arrow@text{#1}}} 
    }{ %
      \arrow@text{#1\;} %
    } %
  } %
}
\makeatother

\newcommand\tdlim{\labeledRightArrow{td}}
\newcommand\rtdlim{\labeledRightArrow{rtd}}

\theoremstyle{plain} 
\newtheorem{prop}{Proposition}

\newtheorem{theorem}{Theorem}

\makeatletter
\newcommand{\theoremtitle}[1]{{\normalfont (#1)}\par\nobreak\@afterheading}
\newcommand{\theoremtitlecite}[2]{{\normalfont (#1~#2)}\par\nobreak\@afterheading}
\makeatother

\theoremstyle{definition}
\newtheorem{defi}[prop]{Definition}
\newtheorem{rmk}[prop]{Remark}

\newtheoremstyle{assumption}
{}
{}
{\normalfont}
{}
{\bfseries}
{}
{ }
{\bfseries\thmnote{#3}}
\theoremstyle{assumption}
\newtheorem*{assumptionthm}{}

\makeatletter
\newlength{\assumptionIndent}
\newenvironment{assumption}[2]{
    \begin{assumptionthm}[\upshape #1]{\normalfont\bfseries #2.}\par\nobreak\@afterheading
        \leftskip\assumptionIndent
        \addtolength{\@rightskip}{\assumptionIndent}
        \addtolength{\rightskip}{\assumptionIndent}
}{
    \end{assumptionthm}
}
\makeatother

\newcommand{\quadtext}[1]{\quad\text{#1}\quad}
\newcommand{\alignindent}{\hspace{-1em}}

\newcommand{\Ldiam}[1]{\mathop{d^{\Lambda_{#1}}{\operatorname{-diam}}}}
\newcommand{\diam}{\operatorname{diam}}

\let\Gammar\Gamma
\renewcommand{\Gamma}{{\mathbb{Z}^d}}

\newcommand{\epsi}{\varepsilon}
\newcommand{\E}{{\mathrm{e}}}
\newcommand{\I}{\mathrm{i}}
\newcommand{\R}{ \mathbb{R} }
\newcommand{\N}{ \mathbb{N} }
\newcommand{\Z}{ \mathbb{Z} }

\newcommand{\unit}{\mathbf{1}}

\newcommand{\Alg}{\mathcal{A}}

\newcommand{\Aloc}{\mathcal{A}_{\mathrm{loc}}}
\newcommand{\D}{\mathrm{d}}
\newcommand{\Or}{{\mathcal{O}}}
\newcommand{\sigmaA}{\sigma_{\mkern-4muA}}

\newcommand{\Lambdak}{{\Lambda_k}} 
\newcommand{\Lambdal}{{\Lambda_l}} 
\newcommand{\LambdaM}{{\Lambda_M}}

\newcommand{\ee}{^{\varepsilon,\eta}}
\newcommand{\eeL}{^{\varepsilon,\eta,\Lambdak}}
\newcommand{\eeLL}{^{\varepsilon,\eta,\Lambdak,\Lambdal}}

\newcommand{\TDL}{\textrm{TDL}}
\newcommand{\RTDL}{\textrm{RTDL}}
\newcommand{\RTDLg}[1]{{\textrm{RTDL}$_#1$}}
\newcommand{\Ab}[1]{\llbracket #1 \rrbracket}

\title{On adiabatic theory for extended fermionic lattice systems}

\author{
    Joscha Henheik%
    \texorpdfstring{%
        \,\orcidlink{0000-0003-1106-327X}
        \footnote{
            \parbox[t]{.7\textwidth}{
                Institute of Science and Technology Austria (IST~Austria),\\
                Am~Campus~1,
                3400~Klosterneuburg,
                Austria\\
                \mbox{Email}:~\href{mailto:joscha.henheik@ist.ac.at}{joscha.henheik@ist.ac.at}\strut
            }%
        }%
    }{}%
    \and%
    Tom Wessel%
    \texorpdfstring{%
        \,\orcidlink{0000-0001-7593-0913}
        \footnote{
            \parbox[t]{.7\textwidth}{
                \foreignlanguage{ngerman}{Fachbereich Mathematik, Eberhard-Karls-Universität~Tübingen,\\
                Auf~der~Morgenstelle~10, 72076~Tübingen,} Germany\\
                \mbox{Email}:~\href{mailto:tom.wessel@uni-tuebingen.de}{tom.wessel@uni-tuebingen.de}
            }%
        }%
    }{}%
}
\begin{document}

\maketitle

\begin{abstract}
    We review recent results on adiabatic theory for ground states of extended gapped fermionic lattice systems under several different assumptions.
    More precisely, we present generalized super-adiabatic theorems for extended but finite and infinite systems, assuming either a \emph{uniform gap} or a \emph{gap in the bulk} above the unperturbed ground state.
    The goal of this note is to provide an overview of these adiabatic theorems and briefly outline the main ideas and techniques required in their proofs.
\end{abstract}

\section{Introduction}
\label{sec:intro}

In this article, we review four recent results on adiabatic theory for ground states of extended finite and infinite fermionic lattice systems at zero temperature~\cite{teufel2020non, henheikteufel20202, henheikteufel20203}.
These results are \emph{generalized super-adiabatic theorems} (see Section~\ref{subsec:NEASS}) and concern Hamiltonians of the form
\begin{equation*}
    H^\epsi = H_0 + \epsi V\,,
\end{equation*}
where the unperturbed Hamiltonian $H_0$ is a sum-of-local-terms (SLT) operator describing short-range interacting fermions and is assumed to have a spectral gap above its ground state.
This gap might be closed by the (small) perturbation $\epsi V$, which is given by a short-range Hamiltonian, a Lipschitz potential, or a sum of both.
Consequently, the results presented in this article are adiabatic theorems for resonances of $H^\epsi$ (cf.~\cite{AF, EH2011}).

The most important corollary and main motivation for proving such theorems in the context of extended fermionic lattice systems is the rigorous justification of linear response theory~\cite{teufel2020non,henheikteufel2020} and the Kubo formula~\cite{kubo} for (topological) insulators~\cite{MPT}, such as quantum Hall systems~\cite{thouless}, where the prototypical relevant perturbation is a linear external potential modeling a constant electric field closing the gap of $H_0$ for every $\epsi \neq 0$ (see Figure~\ref{Intro-fig:intuition local gap} on page~\pageref{Intro-fig:intuition local gap}).

In the remainder of this introduction, we first briefly discuss the connection between linear response and adiabatic theory in Section~\ref{subsec:linearresponse} (see also~\cite{teufel2020non,henheikteufel2020}).
Furthermore, we point out the key ingredients and developments, which allowed to prove the four adiabatic theorems reviewed in this paper.
Afterwards, in Section~\ref{subsec:NEASS}, we explain the notion of {generalized super-adiabatic theorems} and thereby introduce \emph{(super-adiabatic) non-equilibrium almost-stationary states} (NEASSs)~\cite{teufel2020non} as the above mentioned resonances of $H^\epsi$.
A first brief but somewhat precise statement and overview of the results is given in Section~\ref{subsec:statement}.

\subsection{Linear response and adiabatic theory}
\label{subsec:linearresponse}

The formalism of linear response theory~\cite{kubo} has been widely used in physics to calculate the response of a system in thermal equilibrium to external perturbations.
Put briefly, linear response theory provides an answer to the following question: What is the response of a system described by a Hamiltonian $H_0$, that is initially in an equilibrium state $\rho_0$, to a small static perturbation $\epsi V$?
Or, in somewhat more mathematical terms: What is the change%
\footnote{
    To be consistent with the rest of the paper, we view states as linear functionals on the algebra of observables (see Section~\ref{sec:setup}).}
\begin{equation*}
    \rho_\epsi(A) - \rho_0(A) = \epsi \, \sigmaA + o(\epsi)
\end{equation*}
of the expectation value of an observable $A$ induced by the perturbation $\epsi V$ to leading order in its strength $0 <\epsi \ll 1$?
Here, $\rho_\epsi$ denotes the state of the system after the perturbation has been (adiabatically) turned on and $\sigmaA$ denotes the linear response coefficient.

The answer to this fundamental question of linear response clearly hinges on the problem of determining $\rho_\epsi$.
Although in few particular situations one expects $\rho_\epsi$ to remain an equilibrium state for the perturbed Hamiltonian $H^\epsi = H_0 + \epsi V$, the original linear response theory~\cite{kubo} was developed for situations where the system is driven out of equilibrium, i.e.~$\rho_\epsi$ being a resonance state.
As prominently formulated by Simon~\cite{simon1984fifteen} in his `Fifteen problems in mathematical physics' from 1984, the latter non-equilibrium situation causes the main challenges in a rigorous mathematical treatment.
However, in either case, the linear response coefficient $\sigmaA$ is customarily expected to be given by the celebrated \emph{Kubo formula}~\cite{kubo}, and rigorously justifying it was formulated as one of the problems by Simon~\cite{simon1984fifteen}.
For a more detailed recent review on the (mathematical) problem of proving Kubo's formula and its relevance in the context of quantum Hall systems, we refer to~\cite{henheikteufel2020}.

In a nutshell, the problem of justifying linear response theory and proving Kubo's formula is thus to verify that a system, initially in an equilibrium state $\rho_0$, is adiabatically driven by a small perturbation $\epsi V$ into a non-equilibrium state $\rho_\epsi \approx \rho_0$.
Since the perturbation acts over a very long (macroscopic) time, this problem clearly supersedes standard perturbation theory: The change of the state being small is \emph{not} a trivial consequence of the smallness of the perturbation $\epsi V$.
Instead, verifying that the two states, $\rho_\epsi$ and $\rho_0$, are close to each other requires an adiabatic type theorem.

However, even in our rather simple setting (zero temperature, assuming that $\rho_0$ is the gapped ground state of $H_0$ describing an extended fermionic lattice system, the perturbation $\epsi V$ might close the gap), the problem of justifying the linear response formalism also goes beyond standard adiabatic theory.
In fact, the applicability of the standard adiabatic theorem of quantum mechanics is rather restrictive for the following three reasons:
\begin{itemize}
    \item[(i)] The standard adiabatic theorem requires the perturbation $\epsi V$ to \emph{not close the spectral gap}.
        In that scenario, it asserts that $\rho_\epsi$ is (close to) the gapped ground state of $H^\epsi = H_0 + \epsi V$ and as such a (nearly) equilibrium state.
    \item[(ii)] Even if we neglect the first issue, the usual adiabatic theorem estimates the difference between $\rho_\epsi$ and the ground state of the perturbed Hamiltonian $H^\epsi$ in \emph{operator norm}, leaving the translation to local differences in expectation values as an additional and potentially non-trivial step.
    \item[(iii)] In general, extended systems are plagued by the \emph{orthogonality catastrophe}: Whenever for single-particle states $\psi, \tilde{\psi}$ we have ${\Vert \psi - \tilde{\psi} \Vert \sim \epsi}$, the non-interacting many-particle states $\otimes_{x \in \Lambda} \psi_x$ and $\otimes_{x \in \Lambda} \tilde{\psi}_x$ satisfy $\Vert\otimes_{x \in \Lambda} {\psi}_x - \otimes_{x \in \Lambda} \tilde{\psi}_x\Vert \sim \epsi \vert \Lambda \vert $, i.e.~the norm-estimate deteriorates when $\vert \Lambda \vert \to \infty$.
    This means that the approximation error in the standard adiabatic theorem grows with the systems size, and it is thus not applicable for macroscopic systems.
\end{itemize}

A major breakthrough in overcoming these obstacles has recently been achieved by Bachmann, De~Roeck and Fraas~\cite{bachmann2018adiabatic} (see also their introductory lecture notes~\cite{bachmann2018adiabaticLecture}).
They proved the first adiabatic theorem for extended (but finite) lattice systems with short-range interactions, thereby solving the second and third problem in the list above.
More precisely, their result concerns differences in expectation values and provides error estimates, which are uniform in the system size.

For these lattice systems with short-range interactions, well known \emph{Lieb-Robinson bounds}~\cite{LiebRobinson, nachtergaele2017liebrobinson,nachtergaele2019quasi} ensure a finite speed of correlation and prevent build-up of long-range entanglement.
Having Lieb-Robinson bounds at hand allowed Bachmann~et~al.~\cite{bachmann2012automorphic} to prove that the generator of the \emph{spectral flow}, introduced by Hastings and Wen~\cite{hastings2005quasiadiabatic}, is an SLT operator and thus preserves good locality properties.
The general spectral flow technique can then be used to prove automorphic equivalence of two gapped ground states $\rho_0$ and $\rho_1$ of Hamiltonians $H(0)$ and $H(1)$, respectively: Given a smooth path $s \mapsto H(s)$ of (uniformly) gapped SLT Hamiltonians, their ground states are automorphically equivalent (equal up to a conjugation by unitaries) with the generator of the automorphism being an SLT operator~\cite{bachmann2012automorphic}.
This automorphic equivalence allowed Bachmann et~al.~\cite{bachmann2018adiabatic} to prove a \emph{super-adiabatic theorem} (see Section~\ref{subsec:NEASS} for an explanation of this notion) for such systems, however, still requiring the spectral gap not only for $H_0$ but also for $H^\epsi$, i.e.~the gap must remain open.%
\footnote{
    A slight generalization of their result can be found in~\cite{monaco2017adiabatic}, where the authors used an alternative gauge with a time-dependent vector potential for a quantum Hall model.}

The four theorems presented in this article also solved the last remaining problem given under item~(i) in the above list, i.e.~they allow the perturbation $\epsi V$ to close the spectral gap of $H_0$.
The main idea for establishing this generalization is that a spatially local gap should suffice for an adiabatic theorem to hold.
This underlies the space-time adiabatic perturbation theory originally developed for non-interacting fermions by Panati, Spohn and Teufel~\cite{PST2003,PST2003b}, where one utilizes a gap that exists locally in space (and time) but does not exist globally.
It also underlies the recent results by De~Roeck, Elgart and Fraas~\cite{REF2022}, where an adiabatic theorem holds even if the `spectral gap' is filled with eigenvalues, whose corresponding eigenvectors are spatially localized, leaving a gap (with smaller size) locally open.
Finally, this is also the idea behind the Theorems~\ref{thm:existenceofneass3} and especially~\ref{thm:existenceofneass4} presented below, where one still has an adiabatic type theorem although the gap closes at the boundary of the lattices.

Combining the ideas from the space-time adiabatic perturbation theory with the methods invented in~\cite{bachmann2018adiabatic}, the first of the four theorems presented in this article was proven by Teufel~\cite{teufel2020non}.
It concerns extended but finite systems and requires a spectral gap for $H_0$, uniformly in the system size (see Assumption~\nameref{ass:GAPunif}).
The precise statement is formulated in Theorem~\ref{thm:existenceofneassfinite} below.
In order to extend this result from finite lattices to an infinite system, Henheik and Teufel~\cite{henheikteufel20202} adapted ideas from Nachtergaele, Sims and Young~\cite{nachtergaele2019quasi} on controlling the thermodynamic limit of automorphisms with SLT generators.
This result is formulated in Theorem~\ref{thm:existenceofneass2} below.

So far, all mentioned results were obtained under the assumption of a (uniform) spectral gap for the finite systems (which also implies a gap for the infinite system).
However, the recent result on automorphic equivalence with a \emph{gap only in the bulk} (via the GNS construction) by Moon and Ogata~\cite{MO2020}, opened the door for a new class of adiabatic theorems, where the unperturbed Hamiltonian $H_0$ is no longer required to have a uniform spectral gap.
Instead, Theorem~\ref{thm:existenceofneass3}, originally proven by Henheik and Teufel~\cite{henheikteufel20203}, is a result for the infinite volume states and requires a gap in the bulk.
This technically means a gap for the infinite system (cf.~Assumption~\nameref{ass:GAPbulk}) but can be understood as requiring a local gap in the interior of the finite lattices (cf.~Remark~\ref{rmk:gap in bulk}).

Moreover, by employing strong locality estimates from~\cite{nachtergaele2019quasi, MO2020} and assuming fast convergence of ground states, Theorem~\ref{thm:existenceofneass3} can be traced back to extended but finite systems which only have a gap in the bulk.
This was also proven in~\cite{henheikteufel20203} and is formulated in Theorem~\ref{thm:existenceofneass4} below.

\subsection{Non-equilibrium almost-stationary states}
\label{subsec:NEASS}

For the results presented in this paper, we consider time-dependent families
\begin{equation} \label{eq:basicHamiltonian}
    H^\epsi(t) = H_0(t) +\epsi V(t)\,,\quad t\in I \subset \R\,,
\end{equation}
of many-body  Hamiltonians for  lattice fermions in  $\Gammar \subset \Z^d$  with short-range interactions.
Here, $\Gammar$ will either be a finite box $\Lambda$ or the whole of $\Gamma$.
For each \(t\in I\), we denote by \(\rho_0(t)\) the instantaneous ground state of \(H_0(t)\) on the (quasi-local) algebra of observables $\Alg_\Gammar$.
For simplicity of the presentation, we shall assume that the ground state is unique.%
\footnote{
    We refer to the original papers~\cite{teufel2020non, henheikteufel20202, henheikteufel20203} for the most general assumptions.
    However, note that the results from~\cite{henheikteufel20203}, corresponding to our Theorems~\ref{thm:existenceofneass3} and~\ref{thm:existenceofneass4}, are only formulated for a unique ground state, although the underlying result on automorphic equivalence of gapped phases~\cite{MO2020} can easily be generalized to any gapped pure state (see~\cite[Remark~1.4]{MO2020}).
    In general, allowing for a degenerate ground state (or even a gapped spectral patch) requires understanding an enhanced modification of the spectral flow.}
Moreover, we assume that the ground state is separated by a gap from the rest of the spectrum (see Assumptions~\nameref{ass:GAPunif} and~\nameref{ass:GAPbulk} in Section~\ref{sec:result} for the precise formulation).
The perturbation $V(t)$ can be a Hamiltonian with short-range interactions or a possibly unbounded external Lipschitz potential or a sum of both (see Section~\ref{sec:setup} and the Assumptions~\nameref{ass:INT1}--\nameref{ass:INT4} in Section~\ref{sec:result}).

As mentioned above, the main results presented in this article
are so-called \emph{generalized super-adiabatic theorems} for $\rho_0(t)$, which we briefly explain in the following.
For $\epsi=0$, the results are \emph{`standard' super-adiabatic theorems} and establish the existence of super-adiabatic states $\rho_0^\eta(t)$   on $\Alg_\Gammar$ close to $\rho_0(t)$, i.e.\
\begin{equation*}
    \bigl\lvert\rho_0^\eta(t)(A) - \rho_0(t)(A) \bigr\rvert = \Or (\eta)\,,
\end{equation*}
such that the adiabatic time-evolution $\mathfrak{U}^{ \eta}_{t,t_0} $ on $\Alg_\Gammar$ generated by $\frac{1}{\eta} H_0(\cdot)$ intertwines the super-adiabatic states to all orders in the adiabatic parameter $\eta > 0$, i.e.\
\begin{equation}\label{standardadi}
    \bigl\lvert \rho_0^\eta(t_0)(\mathfrak{U}_{t,t_0}^{ \eta} \Ab{A}) - \rho_0^\eta (t)(A) \bigr\rvert
    =\Or(\eta^\infty)
\end{equation}
for all $A$ in a dense subspace $\mathcal{D}\subset \Alg_\Gammar$.
Throughout the entire paper, we shall study our system in the \emph{Heisenberg picture}, meaning that the observable $A$ evolves in time, not the state $\rho_0^\eta(t_0)$ (see also Proposition~\ref{tdlofcauchyinteractions}).
Note that the comparison state $\rho_0^\eta(t)$ does \emph{not} involve any time evolution but simply depends on the Hamiltonian at time $t$ (see Definition~\ref{def:NEASS} for details).
Here and in the following, we write the arguments of (densely defined) linear operators on $\Alg_\Gammar$ inside the brackets~$\Ab{\cdot}$ for better readability.

For $\epsi>0$, the scope of the adiabatic theorem~\eqref{standardadi} extends considerably since the perturbation $\epsi V(t)$ might close the spectral gap and turn the ground state $\rho_0(t)$ of $H_0(t)$ into an instantaneous resonance state $\Pi^{\epsi}(t)$ for $H^\epsi(t)$.
These states have a life-time of order $\Or(\epsi^{-\infty})$ for the dynamics $s\mapsto \E^{\I s \mathcal{L}_{H^\epsi(t)}}$ with $\mathcal{L}_{H^\epsi(t)}\Ab{\cdot} := [ H^\epsi(t), \, \cdot \,  ]$ (formally) denoting the derivation associated to $H^\epsi(t)$.
That is, for all $n\in\N$ and fixed \(t\), it holds that
\begin{equation*}
    \Bigl\lvert   \Pi^{ \epsi }(t) ( \E^{\I s \mathcal{L}_{H^\epsi(t)}}\Ab{A}) - \Pi^{ \epsi} (t)(A)    \Bigr\rvert
    = \Or\Bigl(\epsi^n \, \bigl(1+ |s|^{d+1}\bigl)\Bigr)\,,
\end{equation*}
which is why they were called \emph{non-equilibrium almost-stationary states} (NEASSs) in this context by Teufel~\cite{teufel2020non}.
The generalized super-adiabatic theorems then establish the existence of a \emph{super-adiabatic NEASS} $\Pi^{ \epsi,\eta}(t)$ on $\Alg_\Gammar$ close to $\Pi^{ \epsi }(t)$ such that
the adiabatic time-evolution $\mathfrak{U}^{\epsi, \eta}_{t,t_0}$ generated by $\frac{1}{\eta} H^{\epsi}(\cdot)$   approximately intertwines the super-adiabatic NEASSs in the following sense: for any $n>d$ and  for all $A\in\mathcal{D}\subset \Alg_\Gammar$, we have
\begin{equation}\label{IntroThm}
    \Bigl\lvert\Pi^{ \epsi,\eta}(t_0)(\mathfrak{U}_{t,t_0}^{\epsi, \eta}\Ab{A}) - \Pi^{ \epsi,\eta} (t)(A)\Bigr\rvert
    =\Or\biggl( \eta^{n-d}+\frac{\epsi^{n+1}}{\eta^{d+1}} \biggr)
\end{equation}
uniformly for $t$ in compact sets, which we call a \emph{generalized super-adiabatic theorem}.

\begin{figure}
    \fcapside{
        \caption{%
            Let \(H_0\) be a Hamiltonian with a gapped sector and a gap \(g\).
            Perturbing with a Lipschitz potential \(v(x)=\varepsilon\,x\), the gap gets closed (for large enough lattices).
            But, as indicated in the figure, a local gap persists and an electron at location $x_0$ would either need to overcome the gap (vertical arrow) or tunnel along the distance \(g/\varepsilon\) (horizontal arrow) in order to make a transition from the gapped sector.~\cite{teufel2020non,henheikteufel2020}
        }
        \label{Intro-fig:intuition local gap}
    }{
        \begin{tikzpicture}[scale=1.5, myLine, dist/.style={myLine=1,-latex,shorten >=#1pt}, dist/.default=1]
            \newcommand{\slope}{1.85}
            \coordinate (x0) at (2.5,0.3);
            \newcommand\slopedLineUp[1]{(-1,#1) -- (4,#1+\slope)}
            \newcommand\slopedLineUpText[2]{(-1,#1) -- node[midway,sloped] {#2} (4,#1+\slope)}
            \newcommand\slopedLineDown[1]{(4,#1+\slope) -- (-1,#1)}
            \begin{scope}[myLine=0.5]
                \clip (0,0) rectangle (3,3);
                \draw[fill=gray!30]  \slopedLineUp{0} --  \slopedLineDown{0.5} -- cycle;
                \path \slopedLineUpText{0.25}{\small{gapped sector}};
                \draw[fill=gray!15]  \slopedLineUp{1.4} -- \slopedLineDown{4} -- cycle;
                \path (0,2.8) node[anchor=west] {\small{rest of the spectrum}};
                \path[name path=upper line] \slopedLineUp{1.4};
                \path[name path=lower line] \slopedLineUp{0.5};
                \path[name path=gap distance] (x0) -- +(0,5);
                \path[name intersections={of=upper line and gap distance, by=gu}];
                \path[name intersections={of=lower line and gap distance, by=gl}];
                \path[name path=tunnel distance] (gl) -- +(-5,0);
                \path[name intersections={of=upper line and tunnel distance, by=tl}];
                \fill (gl) circle[radius=1pt];
                \draw[dist] (gl) -- node[right] {\(g\)} (gu);
                \draw[dist=3] (gl) -- node[below,pos=0.6] {\(g/\varepsilon\)} (tl);
            \end{scope}
            \draw[-latex] (0,0.1) -- (0,3.3) node[right] {\(\sigma(H_0)+\varepsilon\,x\)};
            \draw[-latex] (-0.2,0.3) -- (3.3,0.3) node[below] {\(x\)};
            \draw (x0) +(0,0.075) -- +(0,-0.075) node[below] {\(x_0\)};
        \end{tikzpicture}
    }
\end{figure}
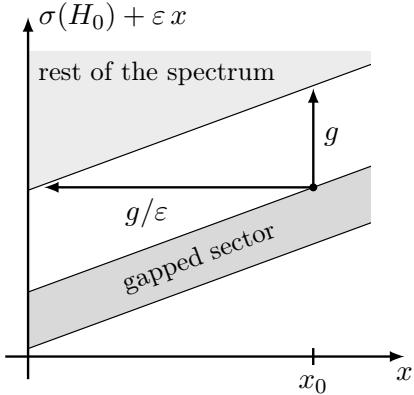

In our setting of gapped Hamiltonians $H_0$ describing insulating materials, there is indeed a clear and simple physical picture suggesting the existence of NEASSs for $H^\epsi$ as observed in~\cite{teufel2020non,henheikteufel2020} (see Figure~\ref{Intro-fig:intuition local gap}).
For simplicity, assume that $H_0$ is a periodic one-body operator in one spatial dimension and that the Fermi energy $\mu$ (chemical potential) lies in a gap of size $g$.
For the perturbation, we consider the potential of a small constant electric field $\epsi$.
In the initial state $\rho_0$, before the perturbation is turned on, all one-body states with energy smaller than~$\mu$ are occupied.
After the voltage has been applied, the energy of an electron located at position~$x_0$ gets substantially shifted by~$\epsi\, x_0$, but is only subject to small force of order~$\epsi$.
As indicated in Figure~\ref{Intro-fig:intuition local gap}, in order to make a transition, such an electron must either overcome the gap of size~$g$ or tunnel a macroscopic distance of order~$g/\epsi$.
Thus, although~$\rho_0$ is neither close to the ground state nor any other equilibrium state of the perturbed Hamiltonian $H^\epsi = H_0 + \epsi V$, it is still almost stationary for~$H^\epsi$.
This heuristic picture remains valid if short-range interactions between the electrons are taken into account.

While for $\epsi=0$ the generalized super-adiabatic theorem~\eqref{IntroThm} reduces to the standard one~\eqref{standardadi},  for $0<\epsi \ll1$ the right-hand side of~\eqref{IntroThm} is small if and only if also $\eta$ is small, but not too small compared to $\epsi$, i.e.\ $\epsi^{\frac{n+1}{d+1}}\ll \eta\ll1$ for some $n\in\N$.
Physically, this simply means that the adiabatic approximation breaks down when the adiabatic switching occurs at times that exceed the lifetime of the NEASS, an effect that has been observed in adiabatic theory for resonances before, see, e.g.,~\cite{AF,EH2011}.
It can also be heuristically understood from the tunneling picture given in Figure~\ref{Intro-fig:intuition local gap}.

Moreover, in view of the linear response problem discussed in Section~\ref{subsec:linearresponse}, let us only mention here that a statement like~\eqref{IntroThm}, in fact, yields a solution to this problem after expanding the state~$\Pi^{\epsi, \eta}(t)$ in powers of~$\epsi$, where the linear term (eventually stemming from the first order operator~$A_1$ given in~\eqref{eq:proof Aj structure}) does, in fact, constitute Kubo's formula.
See~\cite{teufel2020non, henheikteufel2020, henheikteufel20202} for details.

\subsection{Brief statement of the results}
\label{subsec:statement}

We shall establish the existence of super-adiabatic NEASSs in four generally quite different situations, the main differences are also summarized in Table~\ref{tab:classification}:
\begin{itemize}
    \item[(I)] On finite systems \(\Lambda_k \Subset \Gamma\) with suitable boundary conditions, assuming that the unperturbed Hamiltonians \(H_0^{\Lambda_k}(t)\) have a gapped ground state uniformly in \(\Lambda_k\), there exists NEASSs on $\Alg_\Lambdak$ such that the constants in~\eqref{IntroThm} are independent of \(\Lambda_k\).
        See Theorem~\ref{thm:existenceofneassfinite} and~\cite{teufel2020non}.
    \item[(II)] Additionally assuming convergence of the Hamiltonians (they have a thermodynamic limit, cf.~Definition~\ref{def:td}) and ground states, there also exists a super-adiabatic NEASS on $\Alg_\Gamma$ after taking the thermodynamic limit \(\Lambda_k \nearrow \Gamma\).
        See Theorem~\ref{thm:existenceofneass2} and~\cite{henheikteufel20202}.
    \item[(III)] For the infinite system $\Gamma$, assuming that the unperturbed Hamiltonian $H_0$ has a unique gapped ground state (via the GNS construction), there exists a NEASS on $\Alg_\Gamma$, while a (uniform) spectral gap for finite sub-systems is not required.
        See Theorem~\ref{thm:existenceofneass3} and~\cite{henheikteufel20203}.
    \item[(IV)] Additionally assuming a quantitative control on the convergence of the finite volume Hamiltonians \(H^{\Lambda_k}(t)\) (they have a rapid thermodynamic limit, cf.~Definition~\ref{def:rtd}) and the \emph{unperturbed} ground states in the thermodynamic limit, there also exist NEASSs on $\Alg_\Lambdak$ (again with a uniform constant) up to an error vanishing faster than any inverse polynomial in the distance to the boundary.
        See Theorem~\ref{thm:existenceofneass4} and~\cite{henheikteufel20203}.
\end{itemize}

\begin{table}
    \centering
    \caption{Overview of the adiabatic theorems and the original papers.}
    \label{tab:classification}

    \begin{tabular}{l c c}
        \toprule
                        & Finite volume                                                      & Infinite volume                                                     \\
        \midrule
        Uniform gap     & Theorem~\ref{thm:existenceofneassfinite}; see~\cite{teufel2020non} & Theorem~\ref{thm:existenceofneass2}; see~\cite{henheikteufel20202}  \\

        Gap in the bulk & Theorem~\ref{thm:existenceofneass4}; see~\cite{henheikteufel20203} & Theorem~\ref{thm:existenceofneass3}; see~\cite{henheikteufel20203}  \\
        \bottomrule
    \end{tabular}
\end{table}
A typical example of a physically relevant class of Hamiltonians~\cite{monaco2017adiabatic,teufel2020non,henheikteufel2020}, to which the above generalized super-adiabatic theorems apply, is given by
\begin{equation} \label{eq:examplehamiltonian2}
    \begin{aligned}
        H_0^{\Lambda_k}
        ={}& \; \sum_{\mathclap{x,y \in \Lambda_k}} a^*_x \, T(x -y) \, a^{}_y
        \;+\; \sum_{x \in \Lambda_k} a^*_x \, \phi(x) \, a^{}_x
        \\&+ \; \sum_{\mathclap{x,y \in \Lambda_k}} a^*_x a^{}_x \, W\bigl(d^{\Lambda_k}(x,y)\bigr) \, a^*_y a^{}_y
        \;-\; \mu \, N_{\Lambda_k}\,,
    \end{aligned}
\end{equation}
modeling Chern or topological insulators.
In agreement with the precise locality assumptions~\nameref{ass:INT1}--\nameref{ass:INT4} in Section~\ref{sec:result}, we suppose that the kinetic term $T \colon \Gamma \to \mathcal{L}(\mathbb{C}^r)$ is an exponentially  decaying function with $T(-x) = T(x)^*$, the potential term $\phi\colon \Gamma \to \mathcal{L}(\mathbb{C}^r)$ is a bounded function taking values in the self-adjoint matrices, and the two-body interaction $W\colon [0,\infty) \to \mathcal{L}(\mathbb{C}^r)$ is exponentially decaying and also takes values in the self-adjoint matrices.
Note, that $x-y$ in the kinetic term refers to the difference modulo the imposed boundary condition on $\Lambdak$.
In the first line of~\eqref{eq:examplehamiltonian2}, \(a^{}_x\) is the column vector of the annihilation operators \(a^{}_{x,i}\) (\(i\) labels internal degrees of freedom such as spin) and \(a^*_x\) the row vector of the creation operators \(a^*_{x,i}\) (see Section~\ref{sec:setup}).
And with a slight abuse of notation in the second line of~\eqref{eq:examplehamiltonian2}, we wrote $a^*_x a^{}_x$ for the \emph{row} vector with entries $a^*_{x,i} a^{}_{x,i}$ and $a^*_y a^{}_y$ for the \emph{column} vector with entries $a^*_{y,i} a^{}_{y,i}$.

It is well known that non-interacting Hamiltonians $H_0$, i.e.\ with $W \equiv 0$, of the type~\eqref{eq:examplehamiltonian2}   on a \emph{torus} (periodic boundary condition) have a \emph{uniform spectral gap} (see Assumption~\nameref{ass:GAPunif}) whenever the chemical potential $\mu$ multiplying the number operator lies in a gap of the spectrum of the corresponding one-body operator on the infinite domain.
It was recently shown~\cite{hastings2019stability,DS2019}, that  the spectral gap remains open when perturbing by  sufficiently small short-range interactions $W \neq 0$.
On the other hand,  the Hamiltonian $H_0$ on a \emph{cube}  with open boundary condition has, in general, no longer a spectral gap because of the appearance of edge states.
However, away from the boundary, a \emph{gap in the bulk} (see Assumption~\nameref{ass:GAPbulk}) is still present.
While also \emph{uniqueness} of the ground state is expected to hold for such models, to our knowledge it has been shown only for certain types of quantum spin systems, cf.~\cite{yarotsky2005uniqueness, FP2020, nachtergaele2020quasi, nachtergaele2021stability, HTW2022}.
For further details, we refer to the original papers~\cite{teufel2020non, henheikteufel20202, henheikteufel20203}.
Finally, it is an interesting program to extend Table~\ref{tab:classification} by further rows representing different notions of a spectral gap for $H_0$, e.g.~a \emph{local gap} as in~\cite{HTW2022} or even only a \emph{mobility gap} (see~\cite{REF2022} for a first result in this direction).

After a brief introduction to the relevant mathematical framework in Section~\ref{sec:setup}, we formulate the four main theorems in Section~\ref{sec:result}.
Ideas of their proofs are provided in Section~\ref{sec:proof}.

\section{Mathematical Framework}
\label{sec:setup}

In this section, we briefly introduce the (standard) mathematical framework used in the formulation of the adiabatic theorems.
More explanations and details are provided in~\cite{teufel2020non, henheikteufel20202, henheikteufel20203}.

\subsection{Algebra of observables}
We consider fermions with $r$ spin or other internal degrees of freedom on the lattice~$\Gamma$.
Let $\{X \Subset \Gamma \} := \{X \subset \Gamma : \vert X \vert < \infty\}$ denote the set of finite subsets of $\Gamma$, where $\vert X \vert $ is the number of elements in $X$.
For each $X\Subset \Gamma $ let $\mathfrak{F}_X$ be the fermionic Fock space built up from the one-body space $\ell^2(X,\mathbb{C}^r)$.
The  $C^*$-algebra of bounded operators
$\Alg_{X} := \mathcal{L}(\mathfrak{F}_X)$  is generated by the identity element $\mathbf{1}_{\Alg_X}$ and the creation and annihilation operators \(a^*_{x,i}\),~\(a^{}_{x,i}\) for $x \in X$ and $1 \le i \le r$, which satisfy the canonical anti-commutation relations (CAR).
Whenever $X\subset X'$, then $\Alg_{X}$ is naturally embedded as a subalgebra of $\Alg_{X'}$.
For infinite systems, the \emph{algebra of local observables} is defined as the inductive limit
\begin{equation*}
    \Alg_{\mathrm{loc}}  := \; \bigcup_{\mathclap{X  \Subset \Gamma}} \; \Alg_{X}\,,
    \qquad\mbox{and its completion}\qquad
    \Alg_\Gamma := 	\overline{\Alg_{\mathrm{loc}}}^{\Vert \cdot \Vert}
\end{equation*}
with respect to the operator norm $\Vert \cdot \Vert$ is a $C^*$-algebra, called the \emph{quasi-local algebra}.
The even elements $\Alg^+_\Gamma \subset \Alg_\Gamma$ form a $C^*$-subalgebra.
Also, note that for any $X \Subset \Gamma$ the set of elements~$\Alg_{X}^N$ commuting with the number operator $	N_{X} := \sum_{x \in X} a^*_x \, a^{}_x := \sum_{x \in X} \sum_{i=1}^{r}a^*_{x,i}\,a^{}_{x,i}$ forms a subalgebra of the even subalgebra, i.e. $\Alg_X^N \subset \Alg_X^+ \subset \Alg_X$.
As only  even observables will be relevant to our considerations, we will drop the superscript~$^+$ from now on and redefine $\Alg_\Gamma := \Alg^+_\Gamma$.

Since a very similar construction is common for quantum spin systems (see, e.g.,~\cite{nachtergaele2019quasi}),  all the results immediately translate to this setting.

\subsection{Interactions and operator families}
\label{subsec:interactions}

We shall consider sequences of Hamiltonians defined on centered boxes $\Lambda_k := \{-k, \dotsc, +k\}^d$  of size $2k$ with  metric $d^{\Lambda_k}(\cdot,\cdot)$.
This \emph{metric} may differ from the standard $\ell^1$-distance $d(\cdot, \cdot )$ on $\Gamma$ restricted to $\Lambda_k$ if one considers discrete tube or torus geometries,   but satisfies the bulk-compatibility condition
\begin{equation*}
    \forall k \in \mathbb{N} \ \forall x,y \in \Lambda_k: d^{\Lambda_k}(x,y) \le d(x,y) \ \  \text{and} \ \   d^{\Lambda_k}(x,y) = d(x,y) \ \text{whenever} \ d(x,y) \le k \,.
\end{equation*}
An  \emph{interaction on a domain $\Lambda_k$} is a map
\begin{equation*}
    \Phi^{\Lambda_k} \colon \left\{X \subset \Lambda_k\right\} \to  \Alg_{\Lambda_k}^N\,,\;  X \mapsto \Phi^{\Lambda_k} (X) \in \Alg_X^N
\end{equation*}
with values in the self-adjoint operators.
Note that the maps $\Phi^{\Lambda_k}$ can   be extended to maps on the whole $\{X \Subset \Gamma  \}$ or restricted to a smaller $\Lambda_l$, trivially.
In order to describe fermionic systems on the lattice $\Gamma$ in the thermodynamic limit, one considers sequences
$\Phi = \bigl(\Phi^{\Lambda_k}\bigr)_{k \in \mathbb{N}}$ of interactions on domains $\Lambda_k$ and calls the whole sequence an \emph{interaction}.

An \emph{infinite volume interaction} is a map
\begin{equation*}
    \Psi \colon \{ X \Subset \Gamma  \} \to \Alg_\mathrm{loc}^N\,,\;  X \mapsto \Psi (X)\in \Alg_{X}^N \,,
\end{equation*}
again with  values in the self-adjoint operators.
Such an  infinite volume interaction defines a general interaction $\Psi = \big(\Psi^{ \Lambda_k}\big)_{k \in \mathbb{N}}$ by restriction, i.e.\ by setting   $\Psi^{ \Lambda_k} := \Psi|_{\{ X \subset  \Lambda_k\}}$.%
\footnote{
    We will use the convention that $\Phi$ denotes general interactions and $\Psi$ infinite volume interactions.}
With any interaction $\Phi$, one associates an \emph{operator family}, which is a sequence
$A= (A^{\Lambda_k})_{k\in \mathbb{N} }$ of self-adjoint operators
\begin{equation*}
    A^{\Lambda_k} := A^{\Lambda_k}(\Phi) := \sum_{\mathclap{X \subset \Lambda_k}} \Phi^{\Lambda_k} (X) \in \Alg_{\Lambda_k}^N.
\end{equation*}

For any $a > 0$ and $n \in \N_0$, we define the norm
\begin{equation}\label{normdefinition}
    \lVert \Phi \rVert_{a,n }
    := \adjustlimits \sup_{k \in \N} \sup_{x,y \in \Gamma} \sum_{\substack{ X \subset \Lambda_k: \\x,y\in X}} \Ldiam{k}(X)^n \, \E^{a \cdot d^{\Lambda_k}(x,y)}\, \lVert \Phi^{\Lambda_k}(X)\rVert
\end{equation}
on the space of interactions.%
\footnote{
    One should be aware that the norm definition~\eqref{normdefinition} is slightly modified compared to the original works~\cite{teufel2020non, henheikteufel20202, henheikteufel20203} for the sake of simplicity in presentation.
    For more general and precise statements of the theorems we refer the reader to the original works.}
Note that these norms depend on the sequence of metrics $d^{\Lambda_k}$ on the cubes $\Lambda_k$, i.e.\ on the boundary conditions.

Similar constructions for interactions and interaction norms are long known.
More commonly, the norms are independent of the particular lattice \(\Lambda_k\) and the interaction $\big(\Phi^\Lambdak\big)_{k \in \N}$ is given by restrictions of a single infinite volume interaction.
Moreover, in earlier works~\cite{robinson1967,ruelle1969} the authors did not require additional decay properties, which were only added later (see, e.g.,~\cite{simon,HK2006,nachtergaele2019quasi}).
The use of interactions and corresponding norms, which are \emph{not} simply restrictions of an infinite volume interaction, originates in~\cite{monaco2017adiabatic} to incorporate non-trivial boundary conditions.
In order to control commutators with Lipschitz potentials (see~Section~\ref{subsec:lipschitz-potentials}), the dependence on the diameter \(\Ldiam{k}(X)\) was added in~\cite{teufel2020non}.
Finally, to ensure the existence of the thermodynamic limit, it is necessary to require the bulk-compatibility condition~\cite{henheikteufel20202,henheikteufel20203}.
Yet another variant of defining interaction norms is to replace \(\operatorname{dist}(x,y)\) with \(\diam(X)\) in~\eqref{normdefinition} (see, e.g.,~\cite{HK2006,bachmann2021}).

In order to quantify the difference of interactions in the bulk (see Section~\ref{subsec:bulkgap}), we also introduce for any interaction $\Phi^{\Lambda_l}$ on the domain $\Lambda_l$ and any $\Lambda_M\subset \Lambda_l$ the quantity
\begin{equation*}
    \lVert \Phi^{\Lambda_l} \rVert_{a, n,\Lambda_M}
    := \sup_{x,y \in \Lambda_M} \sum_{\substack{ X \subset \Lambda_M: \\x,y\in X}} \diam(X)^n \, \E^{a \cdot d (x,y)} \, \lVert \Phi^{\Lambda_l}(X)\rVert\,,
\end{equation*}
where $d$ and \(\diam\) now refer to the $\ell^1$-distance on $\Gamma$.

Let $\mathcal{B}_{a, n }$ be the \emph{Banach space of interactions} with finite $\lVert \cdot \rVert_{a,n}$-norm and define the space of \emph{exponentially localized interactions} as the intersection $\mathcal{B}_{a, \infty } := \bigcap_{n \in \mathbb{N}_0} \mathcal{B}_{a, n }$.
In the literature, the vector spaces of operator families, which can be written in terms of such interactions, are denoted by $\mathcal{L}_{a,n}$ and $\mathcal{L}_{a,\infty}$.
Moreover, we will be a bit sloppy in the following terminology and call the elements \(A^\Lambdak\) of an operator sequence $A$ \emph{sum-of-local-terms} (SLT) operators, whenever its interaction $\Phi_A$ has a finite interaction norm similar to~\eqref{normdefinition}, but with the exponential replaced by a function growing faster than any polynomial.
This will allow us to formulate the results and the ideas of the proofs without too many details.
For the precise conditions see, e.g.,~\cite[Section~2.2]{henheikteufel20202}.

Now, let $I \subset \mathbb{R}$ be an open interval.
We say that a map $\Phi\colon I \to \mathcal{B}_{a,n}$ is \emph{smooth and bounded} whenever it is (i) term- and point-wise smooth in $t \in I$, i.e. \(t\mapsto\Phi^{\Lambda_k}(t,X)\) are \(C^\infty\)-functions for all \(k\in \N\) and \(X\subset \Lambda_k\), and (ii) $\sup_{t\in I} \Vert \frac{\D^i}{\D t^i} \, \Phi(t)\Vert_{a, n } < \infty$ for all \(i\in \N_0\).
The corresponding spaces of smooth and bounded time-dependent interactions and operator families are denoted by $\mathcal{B}_{I, a, n }$ and $\mathcal{L}_{I, a, n}$ and are equipped with the norm $\Vert\Phi \Vert_{I, a, n } := \sup_{t \in I} \Vert \Phi (t)\Vert_{a, n }$.
We say that $\Phi\colon I \to \mathcal{B}_{a, \infty }$ is smooth and bounded, if $\Phi\colon I \to \mathcal{B}_{a,n }$ is smooth and bounded for all $n \in \N_0$, and we write $\mathcal{B}_{I, a, \infty }$ and~$\mathcal{L}_{I, a, \infty }$ for the corresponding spaces of \emph{time-dependent exponentially localized interactions} and \emph{operator families} respectively.

For (time-dependent) \emph{infinite volume interactions} $\Psi$, we add a superscript~$^\circ$  to the norms and to the normed spaces defined above, emphasizing in particular the use of open boundary conditions, i.e.\
$d^{\Lambda_k} \equiv d$.
Note that the compatibility condition for the metrics $d^{\Lambda_k}$ implies that $\Vert \Psi \Vert_{a, n } \le \Vert \Psi \Vert_{a, n }^\circ $.

\subsection{Lipschitz potentials}
\label{subsec:lipschitz-potentials}

For the  perturbation we will allow external potentials
$
    v = \left( v^{\Lambda_k}\colon \Lambda_k \to \mathbb{R}\right)_{ k \in \mathbb{N}}
$
that satisfy the Lipschitz condition
\begin{equation} \label{eq:Lipcond}
    \mathcal{C}_v := \sup_{k \in \mathbb{N}} \,\sup_{\substack{x,y \in \Lambda_k:\\x\neq y}} \frac{\bigl\lvert v^{\Lambda_k}(x)- v^{\Lambda_k}(y)\bigr\rvert}{ d^{\Lambda_k}(x,y)} < \infty\,,
\end{equation}
and call them for short \emph{Lipschitz potentials}.%
\footnote{
    Teufel~\cite{teufel2020non} instead allowed slightly more general \emph{slowly-varying potentials}.
    And while the phrase captures the idea very well, the technical definition is less transparent and slightly complicates the presentation of the proofs.
    Hence, we here, as in~\cite{henheikteufel20202,henheikteufel20203}, restrict to the subclass of Lipschitz potentials.
}
With a Lipschitz potential $v$ we associate the corresponding operator-sequence $V_v = (V_v^{ \Lambda_k})_{ k \in \mathbb{N}}$ defined by
\begin{equation*}
    V_v^{\Lambda_k} := \sum_{x \in \Lambda_k} v^{ \Lambda_k}(x) \, a^*_x \, a^{}_x
\end{equation*}
and denote the space of Lipschitz potentials by $\mathcal{V}$.
We emphasize that, since \(\sup_{k \in \mathbb{N}} \sup_{\substack{x\in \Lambda_k}} \lvert v^{\Lambda_k}(x) \rvert\) might be infinite, \(V_v\) is in general no SLT~operator.
However, this is still more restrictive than general onsite potentials, because it only varies slowly in space.
Moreover, we say that the map $v\colon I \to \mathcal{V}$ is smooth and bounded whenever (i)~$v^\Lambdak(x, \cdot)$ are \(C^\infty\)-functions for all $k \in \N$ and $x \in \Lambdak$, and (ii)~satisfies $\sup_{t \in I}C_{ \frac{\D^i}{\D t^i}v (t)} < \infty$ for all $i \in \mathbb{N}_0$.
The space of smooth and bounded \emph{time-dependent Lipschitz potentials} is denoted by $\mathcal{V}_I$.

As above, we also introduce \emph{infinite volume Lipschitz potentials} $v_\infty\colon\Gamma\to \R$, which, again by restriction and invoking the compatibility condition for the metrics $d^{\Lambda_k}$, can be viewed as a Lipschitz potential with $d^\Lambdak \equiv d$ in~\eqref{eq:Lipcond}.
And analogously to Section~\ref{subsec:interactions}, for (time-dependent) \emph{infinite volume Lipschitz potentials}, we add a superscript~$^\circ$ to the constant from~\eqref{eq:Lipcond} and to the spaces, emphasizing the use of open boundary conditions.
Note that the compatibility condition for the metrics $d^{\Lambda_k}$ implies that $\mathcal{C}_v \ge \mathcal{C}_v^\circ $.

\section{Adiabatic theorems for gapped quantum systems}
\label{sec:result}

As mentioned in the introduction, we shall distinguish two generally quite different settings regarding the presence of a spectral gap of the unperturbed Hamiltonian $H_0$ grouped as Theorem~\ref{thm:existenceofneassfinite} and Theorem~\ref{thm:existenceofneass2} in Section~\ref{subsec:unifgap} as well as Theorem~\ref{thm:existenceofneass3} and Theorem~\ref{thm:existenceofneass4} in Section~\ref{subsec:bulkgap}.
First, in Section~\ref{subsec:unifgap}, we will work under the assumption that there exists a sequence of subsystems $(\Lambda_k)_{k \in \N}$ equipped with an appropriate metric (reflecting, e.g., periodic boundary conditions), ensuring that $H_0^{\Lambda_k}$ have a \emph{uniform gap} above their ground state, which is made precise in Assumption~\nameref{ass:GAPunif} below.
Then, in Section~\ref{subsec:bulkgap}, however, we drop this assumption and solely assume that $H_0$ has a \emph{gap in the bulk}, meaning that the GNS Hamiltonian, describing the system in the thermodynamic limit, has a spectral gap above its ground state eigenvalue zero (see Assumption~\nameref{ass:GAPbulk}).
Note that the second group of results is more general than the first group with regard to the gap condition, since a uniform gap for finite systems guarantees a spectral gap for the GNS Hamiltonian describing the infinite system (see Proposition~5.4 in~\cite{BDN2015}).
Therefore, the second row in Table~\ref{tab:classification} somewhat improves the results in the first row since finding a suitable geometry for which one already has a spectral gap for finite systems is no longer necessary.

In the precise formulation of the adiabatic theorems, we shall frequently use the abbreviating phrase that a state $\Pi^{\epsi, \eta}(t)$ \emph{is a super-adiabatic NEASS} (see Section~\ref{subsec:NEASS}), which we generally define as follows, reminiscent of~\cite{teufel2020non,henheikteufel20202,henheikteufel20203}.

\begin{defi}
    \theoremtitle{Super-adiabatic non-equilibrium almost-stationary states}
    \label{def:NEASS}
    We assume to be in the following general setting, which is made precise in concrete situations: For (small) $\epsi > 0$, define the time-dependent Hamiltonian
    \begin{equation*}
        H^\epsi(t) =  H_0 (t) + \epsi V(t)\,, \ t \in I\,, \quad \text{on} \quad \Gammar \subset \Gamma
    \end{equation*}
    and let $\rho_0(t)$ be (close to)%
    \footnote{
        See the comment below Assumption~\nameref{ass:Sbulk} on page~\pageref{ass:Sbulk} for a precise definition.
    }
    the ground state of $H_0(t)$.
    Moreover, denote the Heisenberg time-evolution on the algebra of (quasi-local) observables $\Alg_\Gammar$ generated by $\frac{1}{\eta} H^\epsi(t)$ as $\mathfrak{U}_{t,t_0}^{\epsi, \eta}$, where $t, t_0 \in I $ for some open interval $I \subset \R$ and $\eta > 0$ is a (small) adiabatic parameter.

    Then, we say that a state $\Pi^{\epsi, \eta}(t)$ on $\Alg_\Gammar$ \emph{is a super-adiabatic non-equilibrium almost-stationary state} for the state $\rho_0(t)$ and the time-evolution $\mathfrak{U}_{t,t_0}^{\epsi, \eta}$ on $\Alg_\Gammar$ if it satisfies the following properties:
    \begin{enumerate}
        \item $\Pi^{\varepsilon, \eta}$ almost {\bf intertwines the   time evolution}: For any {$n \in \mathbb{N}$}, there exists a constant $C_n$ such that for any $ t,t_0 \in I$ and for all $X \Subset \Gammar$ and $A \in \Alg_X \subset \Alg_\Gammar$ we have
              \begin{equation} \label{eq:generalbound}
                  \Bigl\lvert \Pi^{\varepsilon, \eta}(t_0)(\mathfrak{U}_{t,t_0}^{\varepsilon, \eta}\Ab{A}) - \Pi^{\varepsilon, \eta}(t)(A) \Bigr\rvert   \le \ C_n \ \frac{\varepsilon^{n+1} + \eta^{n+1}}{\eta^{d+1}}   \ \left(1+\vert t- t_0\vert^{d+1}\right) \,  \Vert A \Vert \, \vert X \vert^2.
              \end{equation}
              \label{def:NEASS-general-bound}
        \item $\Pi^{\varepsilon, \eta}$  is  {\bf local in time}:
              \(\Pi^{\varepsilon,\eta}(t)\) only depends on ${H_0}$ and \(V\) and their time derivatives at time~$t$.
              \label{def:NEASS-local-in-time}
        \item $\Pi^{\varepsilon, \eta}$  is \textbf{stationary} whenever the Hamiltonian is stationary:
              If for some fixed $t\in I$ all time-derivatives of $H_0$ and $V$ vanish at time $t$,
              then $\Pi^{\varepsilon, \eta}({t})$ equals the NEASS%
              \footnote{
                  It follows from the construction sketched in Section~\ref{sec:proof-finite-systems-uniform-gap} that \(\Pi^{\varepsilon}({t})=\Pi ^{\varepsilon,0}({t})\).
                  Moreover, \(\Pi^{\varepsilon}({t})\) is almost stationary with a bound as in~\eqref{eq:generalbound} where the fraction is replaced by \(\varepsilon^{n+1}\).
              }%
              ~\(\Pi ^{\varepsilon}({t})\)
              for the instantaneous ground state \(\rho_0(t)\) and the time-evolution \(s\mapsto\E^{\I s\mathcal{L}_{H^\varepsilon(t)}}\) generated by the time-independent Hamiltonian \(H^\varepsilon(t)\).
              \label{def:NEASS-stationary}
        \item $\Pi^{\varepsilon, \eta} $  equals the (approximate) ground state $\rho_0$  of $H_0$ whenever the perturbation vanishes and the Hamiltonian is stationary:
              If for some $t\in I$ all time-derivatives of $H_0$ and \(V\) vanish at time $t$ and $V(t) = 0$,
              then $\Pi^{\varepsilon, \eta}(t) = \Pi^{\varepsilon, 0}(t) = \rho_0(t)$.
              \label{def:NEASS-equals-ground-state}
    \end{enumerate}
\end{defi}

We could have written bound~\eqref{eq:generalbound} in a more general form as indicated by~\eqref{IntroThm}.
For example, we could allow $(1 + |t-t_0|^{d+1})$ to be replaced by a constant $C_K < \infty$, depending only on a compact subset $K \subset I$ of times, or, similarly, $|X|^2$ to be replaced by a constant $C_X < \infty$, depending only on the support $X \Subset \Gamma$ of the observable $A$.
Also, the power of $\eta$ in the denominator could be allowed to be more general, e.g.~some constant $C_d < \infty$ instead of $d + 1$.
However, the concrete form of~\eqref{eq:generalbound} indeed matches the precise bounds of the results in Section~\ref{sec:result}.

\subsection{Systems with a uniform gap}
\label{subsec:unifgap}

Throughout this section, we assume that $H_0$ has a uniformly gapped unique ground state in the following sense.

\begin{assumption}{(GAP\textsubscript{unif})}{Assumptions on the ground state of $H_0$}\label{ass:GAPunif}\noindent
    Let $\Phi_{H_0} = \big(\Phi_{H_0}^{\Lambdak}\big)_{k \in \N}$ be an interaction.
    There exists   $L \in \mathbb{N}$ such that for all $t \in I$, $k \ge L$ and corresponding $\Lambdak$ the operator $H_0^{\Lambdak}(t)$ has a simple gapped ground state eigenvalue $E_0^{\Lambdak}(t) = \inf \sigma(H_0^{\Lambdak}(t))$, i.e.~there exists $g> 0$ such that $\operatorname{dist}\bigl(E_0^{\Lambdak}(t), {\sigma(H_0^{\Lambdak}(t)) \setminus  \{E_0^{\Lambdak}(t)\}}\bigr) \ge g$, for all $t \in I$, $k \ge L$.
    We denote the spectral projection of $H_0^{\Lambdak}(t)$ corresponding to  $E_0^{\Lambdak}(t) $ by $P_0^{\Lambdak}(t)$ and write $\rho_0^{\Lambdak}(t)(\cdot ) := \operatorname{tr} \big(P_0^{\Lambdak}(t) \, \cdot \, \big)$ for the canonically associated state on $\Alg_\Lambdak$.
\end{assumption}

A physically relevant class of Hamiltonians satisfying this assumption (possibly up to the uniqueness, which we require for simplicity of the presentation) was given in~\eqref{eq:examplehamiltonian2} in Section~\ref{subsec:statement}.
In the following, we shall present adiabatic theorems for extended but finite systems (Theorem~\ref{thm:existenceofneassfinite}) as well as for infinite systems (Theorem~\ref{thm:existenceofneass2}) under Assumption~\nameref{ass:GAPunif}.

\subsubsection{Extended but finite systems}
The basic assumption on the Hamiltonian says that it is composed of exponentially localized interactions and/or a Lipschitz potential.

\begin{assumption}{(INT\textsubscript{1})}{Assumptions on the interactions}
    \label{ass:INT1}\noindent
    Let $H_0, H_1$ be the Hamiltonians of two time-dependent exponentially localized interactions, i.e.~$\Phi_{H_0}, \Phi_{H_1} \in  \mathcal{B}_{I, a,\infty }$ for some $a>0$, and $v\in \mathcal{V}_I$ be a time-dependent Lipschitz potential.
\end{assumption}

The following results due to Teufel~\cite{teufel2020non} marks the starting point for generalized super-adiabatic theorems for extended fermionic lattice systems.

\begin{theorem}
    \theoremtitlecite{Adiabatic theorem for finite systems with a uniform gap}{\cite[Theorem~5.1]{teufel2020non}}
    \label{thm:existenceofneassfinite}
    Under Assumptions~\nameref{ass:GAPunif} and~\nameref{ass:INT1}, there exists a sequence of near-identity%
    \footnote{
        Indeed, \(\sup_{k\in \N} \bigl\lVert A - \beta^{\varepsilon,\eta,\Lambda_k}(t) \Ab{A} \bigr\rVert \leq (\varepsilon+\eta) \, C \, \lVert A \rVert \, \lvert X \rvert \, \lvert t \rvert\) for \(A\in \Alg_X\) and \(t\) in a bounded interval by~\cite[Theorem~3.4(i)]{nachtergaele2019quasi}.
    }
    automorphisms $\beta^{\varepsilon, \eta, \Lambdak}(t)= \E^{\I \varepsilon\mathcal{L}^\Lambdak_{S^{\varepsilon, \eta}(t)}}$ with SLT generators $ S^{\varepsilon, \eta}$ for any $\varepsilon, \eta \in (0,1]$ and $t \in I$ such that the states
    \begin{equation} \label{eq:NEASSfinite}
        \Pi^{\varepsilon, \eta, \Lambdak}(t) := \rho_0^\Lambdak(t)\circ \beta^{\varepsilon, \eta, \Lambdak}(t)
    \end{equation}
    are super-adiabatic NEASSs for the Heisenberg time-evolution $\mathfrak{U}_{t,t_0}^{\epsi, \eta, \Lambdak}$ on $\Alg_\Lambdak$ generated by $\tfrac{1}{\eta} \, H^{\epsi, \Lambdak}(\cdot)$ with
    \begin{equation*}
        \tfrac{1}{\eta} \, H^{\varepsilon,\Lambda_k}(t)
        := \tfrac{1}{\eta} \,\Bigl( H^{ \Lambda_k}_0(t) + \epsi \,\bigl(V_v^{\Lambda_k} (t) + H^{\Lambda_k}_1(t) \bigr)\Bigr)
    \end{equation*}
    uniformly in $k \ge L$.
    That is, for every $n \in \N$, there exists a constant $C_n$, such that for any   $A \in \Alg_{X}$, $\epsi, \eta \in (0,1]$ and all $t, t_0 \in I$ it holds that
    \begin{equation*}
        \sup_{ k\ge L}   \Bigl\lvert \Pi^{\varepsilon, \eta, \Lambdak}(t_0)(\mathfrak{U}_{t,t_0}^{\varepsilon, \eta, \Lambdak}\Ab{A}) - \Pi^{\varepsilon, \eta, \Lambdak}(t)(A) \Bigr\rvert
        \le
        C_n \, \frac{\varepsilon^{n+1} +  \eta^{n+1}}{\eta^{d+1}} \, \Bigl(1+\vert t- t_0 \vert^{d+1}\Bigr)  \,\lVert A \rVert \, \lvert X \rvert^2 \,.
    \end{equation*}
\end{theorem}

The proof of this result fundamentally builds on space-time adiabatic perturbation theory~\cite{PST2003,PST2003b} and technical estimates originally derived in~\cite{bachmann2018adiabatic}.
The latter show that the operations necessary for the construction of the generator of the near-identity automorphism in the definition of the NEASS in~\eqref{eq:NEASSfinite} (almost) preserve exponential localization required for the Hamiltonian (see Section~\ref{sec:proof}).
As already mentioned in the introduction, although the adiabatic theorem in~\cite{bachmann2018adiabatic} is at first sight quite similar to the one above, it requires the perturbation to \emph{not} close the spectral gap of the Hamiltonian $H_0$ and is thus not generalized in the sense explained in Section~\ref{subsec:NEASS}.

\subsubsection{Infinite systems}
The next result is obtained from Theorem~\ref{thm:existenceofneassfinite} by taking $\Lambdak \nearrow \Gamma$.
This requires the interactions and the Lipschitz potential composing the Hamiltonian~\eqref{eq:basicHamiltonian} to \emph{have a thermodynamic limit}~\cite{henheikteufel20202} in the following sense.

\begin{defi}
    \theoremtitle{Thermodynamic limit of interactions and potentials}
    \label{def:td}
    \vspace{\topsep}
    \begin{itemize}
        \item[(a)] An exponentially localized time-dependent interaction $\Phi  \in \mathcal{B}_{I,a, \infty}$ is said to \textit{have a thermodynamic limit} (have a TDL)
            if there exists an infinite volume interaction $ \Psi \in \mathcal{B}_{I,a, \infty}^\circ$
            such that
            \begin{align*} \hspace{-.5cm}
                \forall \, n \in \N, \, i \in \mathbb{N}_0, \, M \in \mathbb{N} : \; \adjustlimits \lim_{k\to\infty}\sup_{t \in I}\left\Vert \frac{\D^i}{\D t^i}\left( { \Psi}- \Phi^{ \Lambdak}\right) (t)  \right\Vert_{a, n,\LambdaM} =0\,,
            \end{align*}
            and we write $\Phi\tdlim \Psi$ in this case.

            An operator family is said to \textit{have a TDL} if and only if the corresponding interaction does.

            For more general (non-exponentially localized) SLT operators, the definition is completely analogous.
        \item[(b)] 	A Lipschitz potential
            $v \in \mathcal{V}_I$
            is said to \textit{have a TDL} if there exists an infinite volume Lipschitz potential $v_\infty \in \mathcal{V}_I^\circ$   such that
            \begin{align*}
                \forall M \in \mathbb{N} \quad \exists K\ge M  \quad \forall k  \ge K,\,  t\in I: v^{ \Lambdak}(t, \cdot) |_{\LambdaM} = v_{\infty}(t, \cdot) |_{\LambdaM}\,.
            \end{align*}
            Again, we write $v\tdlim v_\infty$ in this case.
    \end{itemize}
\end{defi}

Note that, whenever $\Phi = \Psi$ for some infinite-volume interaction $\Psi$, or $v = v_\infty$ for some infinite volume Lipschitz potential $v_\infty$, both $\Phi$ and $v$ trivially have a \TDL{}.

The following proposition is a standard consequence of Lieb-Robinson bounds and shows that the property of {having a \TDL} for interactions and Lipschitz potentials guarantees the existence of the thermodynamic limit for the corresponding evolution operators~\cite{nachtergaele2019quasi, BdSP}.
We remark that it remains true under less restrictive assumptions on the localization quality of the interaction (see, e.g., Proposition~2.2 in~\cite{henheikteufel20202}).

\begin{prop}
    \theoremtitle{Thermodynamic limit of evolution operators}
    \label{tdlofcauchyinteractions}
    Let $K_0 \in \mathcal{L}_{I, a, \infty }$ and $w \in \mathcal{V}_I$ both have a thermodynamic limit, i.e.\ $\Phi_{K_0}\tdlim \Psi_{K_0} $ and $w\tdlim w_\infty$ for some \(\Psi_{K_0}\in\mathcal{B}_{I,a,\infty}^\circ\) and \(w_\infty\in\mathcal{V}_I^\circ\).
    Set $K = K_0 + V_w$
    and  let $U^{\eta,\Lambdak}(t,t_0)$ denote the evolution family generated by $K^{\Lambdak} (t)$ in scaled time with $\eta >0$, i.e.~the solution to the Schrödinger equation
    \begin{equation*}
        \mathrm{i }\,\eta\,\frac{\mathrm{d}}{\mathrm{d}t}\, U^{\eta,\Lambdak}(t,t_0) = K^{\Lambdak} (t)\, U^{  \eta,\Lambdak}(t,t_0)
    \end{equation*}
    with $U^{\eta,\Lambdak}(t_0,t_0) = \mathrm{id}$.
    Then, there exists a co-cycle of automorphisms $\mathfrak{U}_{t,t_0}^{\eta}:\Alg_\Gamma \to \Alg_\Gamma$ such that for all $A \in \Aloc $,
    \begin{equation*}
        \mathfrak{U}^{\eta}_{t,t_0}\Ab{A} = \lim\limits_{k \to \infty}  \mathfrak{U}^{\eta,\Lambdak}_{t,t_0}\Ab{A}  := \lim\limits_{k \to \infty} U^{\eta,\Lambdak}(t,t_0)^*\, A \,U^{\eta,\Lambdak}(t,t_0)\,.
    \end{equation*}
    The co-cycle  $\mathfrak{U}^{\eta}_{t,t_0}$ only depends on $\Psi_{K_0}$ and $w_\infty$
    and  is generated by the time-dependent (closed) derivation  $ (\mathcal{L}_{K(t)}, D(\mathcal{L}_{K(t)}))$ associated with $ K(t)$.
\end{prop}

As mentioned above, since the following Theorem~\ref{thm:existenceofneass2} is deduced from Theorem~\ref{thm:existenceofneassfinite} by taking $\Lambdak \nearrow \Gamma$, we will need to assume the existence of a thermodynamic limit for the building blocks of the Hamiltonian~\eqref{eq:basicHamiltonian}.

\begin{assumption}{(INT\textsubscript{2})}{Assumptions on the interactions}
    \label{ass:INT2}\noindent
    For $ \Psi_{H_0}, \Psi_{H_1} \in  \mathcal{B}^\circ_{I,a,\infty }$ for some $a >0$ and   $v_\infty \in \mathcal{V}_I^\circ$ there exist $\Phi_{H_0}, \Phi_{H_1} \in \mathcal{B}_{I,a,\infty }$ and $v \in \mathcal{V}_I$
    with appropriate boundary conditions (encoded in the definition of the norms defining the spaces $ \mathcal{L}$ and the Lipschitz condition) all having a \TDL~with the respective object as the limit, i.e.~$\Phi_{H_0}\tdlim \Psi_{H_0} $, $\Phi_{H_1}\tdlim \Psi_{H_1} $ and $v \tdlim v_\infty $.
\end{assumption}

We also assume the convergence of ground states, by means of the Banach-Alaoglu Theorem (the unit sphere in $\Alg_\Gamma^*$ is weak$^*$-compact), essentially only in order to avoid the extraction of a subsequence.

\begin{assumption}{(S\textsubscript{unif})}{Assumptions on the convergence of states}\label{ass:Sunif}\noindent
    Assume that for every $t \in I$ the sequence $\big(\rho_0^{\Lambdak}(t)\big)_{k \in \mathbb{N}}$  of ground states (naturally extended to the whole of $\Alg_\Gamma$) converges in the weak$^*$-topology to a state $\rho_0(t)$ on $\Alg_\Gamma$, which we call the \textit{gapped limit ground state at $t \in I$}.
\end{assumption}

We can now formulate the second generalized super-adiabatic theorem concerning infinite systems with a uniform gap~\cite{henheikteufel20202}.

\begin{theorem}
    \theoremtitlecite{Adiabatic theorem for infinite systems with a uniform gap}{\cite[Theorems~3.2~and~3.5]{henheikteufel20202}}
    \label{thm:existenceofneass2}
    Under the Assumptions~\nameref{ass:GAPunif}, \nameref{ass:INT2} and~\nameref{ass:Sunif}, there exists a near-identity automorphism $\beta^{\varepsilon, \eta}(t)= \E^{\I \varepsilon\mathcal{L}_{S^{\varepsilon, \eta}(t)}}$ with SLT generators $ S^{\varepsilon, \eta}$ for any $\varepsilon, \eta \in (0,1]$ and $t \in I$, such that the state
    \begin{equation*}
        \Pi^{\varepsilon, \eta}(t) := \rho_0(t)\circ \beta^{\varepsilon, \eta}(t)
    \end{equation*}
    is a super-adiabatic NEASS for the Heisenberg time-evolution on $\Alg_\Gamma$ generated by
    $\frac{1}{\eta}\Psi_{H^{\varepsilon}(\cdot)}$ with
    \begin{equation*}
        \Psi_{H^{\varepsilon}}:= \Psi_{H_0} + \epsi \, (V_{v_\infty} + \Psi_{H_1})\,.
    \end{equation*}
\end{theorem}

The crucial point in the proof of Theorem~\ref{thm:existenceofneass2} in~\cite{henheikteufel20202} is to show that the property of {having a TDL} is designed in such a way that it is preserved under all necessary operations for the construction of the NEASS (see Section~\ref{sec:proof}).
Therefore, also the near-identity automorphism from~\eqref{eq:NEASSfinite} converges as $\Lambdak \nearrow \Gamma$ by means of Proposition~\ref{tdlofcauchyinteractions}.

\subsection{Systems with a gap in the bulk}
\label{subsec:bulkgap}

In this section, we drop Assumption~\nameref{ass:GAPunif} of a uniform gap for finite systems, but merely work under the condition of a \emph{gap in the bulk}, which is formulated via the Gelfand-Naimark-Segal (GNS) construction in Assumption~\nameref{ass:GAPbulk} below: Let $\Psi_{H_0}\in \mathcal{B}_{a,0 }^\circ$ be an infinite volume interaction  and $\mathcal{L}_{H_0}$ denote the induced derivation on (a dense subset of) $\Alg_\Gamma$.
A state $\omega$ on $\Alg_\Gamma$ is called an $ \mathcal{L}_{H_0}$-ground state, if and only if $\omega(A^*\mathcal{L}_{H_0}(A)) \ge 0$ for all $A \in D(\mathcal{L}_{H_0})$.
Let $\omega$ be an $\mathcal{L}_{H_0}$-ground state and $(\mathcal{H}_{\omega}, \pi_{\omega}, \Omega_{\omega})$ be the corresponding GNS triple ($\mathcal{H}_{\omega}$ a Hilbert space, $\pi_{\omega}\colon \Alg \to \mathcal{L}(\mathcal{H}_{\omega})$ a representation and $\Omega_{\omega} \in \mathcal{H}_{\omega}$ a cyclic vector).
Then, there exists a unique densely defined, self-adjoint positive operator $H_{0,\omega} \ge 0$ on $\mathcal{H}_{\omega}$ satisfying
\begin{equation}
    \pi_{\omega}(\E^{\I t\mathcal{L}_{H_0}}\Ab{A})
    = \E^{\I tH_{0,\omega}} \, \pi_{\omega}(A) \,\E^{- \I tH_{0,\omega}}
    \quad \text{and} \quad
    \E^{- \I tH_{0,\omega}} \,\Omega_{\omega} = \Omega_{\omega}
\end{equation}
for all $A \in \Alg$ and $t \in \mathbb{R}$.
We call this $H_{0,\omega}$ the \emph{bulk Hamiltonian} (or \emph{GNS Hamiltonian}) associated with $\Psi_{H_0}$ and $\omega$.
See~\cite{BR II 1996} for the general theory.

We assume that $\Psi_{H_0}$ has a unique gapped ground state in the following sense (cf.~\cite{MO2020, henheikteufel20203}):

\begin{assumption}{(GAP\textsubscript{bulk})}{Assumptions on the ground state of $\Psi_{H_0}$}\label{ass:GAPbulk}
    \vspace{\topsep}
    \begin{itemize}
        \item[(i)] \textbf{Uniqueness.}
            For each $t \in I$,  there exists a unique $\mathcal{L}_{H_0(t)}$-ground state~$\rho_0(t)$.
        \item[(ii)] \textbf{Gap.}
            There exists $g>0$ such that $\sigma\bigl(H_{0,\rho_0(t)}(t)\bigr) \setminus \{0\} \subset [g,\infty)$ for all~$t \in I$.
        \item[(iii)] \textbf{Regularity.}
            For any strictly positive $f \in \mathcal{S}(\R)$ (Schwarz functions), define $\mathcal{D}_f$ as the set of observables $A \in \Alg_\Gamma$ for which $\lVert A \rVert_f := \lVert A \rVert + \sup_{k \in \mathbb{N}} \bigl(\lVert (1 - \mathbb{E}_{\Lambda_k})\Ab{A}\rVert / f(k)\bigr) <\infty$, where $\mathbb{E}_{\Lambdak}\Ab{\cdot}$ denotes the conditional expectation (see~\cite[Appendix~C]{henheikteufel20202}).
            Then, for any $A \in \mathcal{D}_f$, $t\mapsto \rho_0(t)(A)$ is differentiable and there exists a constant $C_f$ such that
            \begin{equation*}
                \sup_{t \in I} \vert \dot{\rho_0}(t)(A)\vert \le C_f \, \Vert A\Vert_f\,.
            \end{equation*}
    \end{itemize}
\end{assumption}

The smoothness of expectation values of (almost) exponentially localized observables as under item (iii) is a rather technical condition and a consequence of a uniform gap as in Assumption~\nameref{ass:GAPunif} (see Remark~4.15 in~\cite{MO2020} {and Lemma~6.0.1 in~\cite{moondiss}}).
Although uniqueness of the ground state in item (i), which we required throughout the paper, is expected to hold for the physically relevant type of Hamiltonian~\eqref{eq:examplehamiltonian2}, it has been shown, to our present knowledge, only in very specific quantum spin systems.
These include (a) weak perturbations of non-interacting gapped frustration-free systems~\cite{yarotsky2005uniqueness, HTW2022}, and (b) short-range interacting frustration-free models fulfilling \emph{local topological quantum order} (LTQO)~\cite{nachtergaele2021stability, bachmann2021}.

\begin{rmk} \label{rmk:gap in bulk}
    As mentioned in the beginning of Section~\ref{sec:result}, item (ii) holds, in particular, if one has a uniform gap for finite systems as spelled out in Assumption~\nameref{ass:GAPunif}, since it cannot close abruptly in the thermodynamic limit for the GNS Hamiltonian (see Proposition~5.4 in~\cite{BDN2015}).
    However, we observe that a considerably \emph{weaker} sufficient condition for having a gap for the GNS Hamiltonian as in Assumption~\nameref{ass:GAPbulk}~(ii) is to have a \emph{gap in the bulk for the finite systems} $\Lambdak$ in the following sense: There exists $g > 0$ such that for all $k \in \N$ there exists some $l = l(k) \in \N$ with $l(k) \to \infty$ as $k \to \infty$, and we have
    \begin{equation} \label{eq:finitegapinbulk}
        \rho_0^\Lambdak(t)\bigl(A^* \mathcal{L}_{H_0(t)}^{\Lambdak} \Ab{A}\bigr)
        \ge
        g \, \Bigl(\rho_0^\Lambdak(t)(A^* A) - \bigl\lvert\rho_0^\Lambdak(t)(A)\bigr\rvert^2\Bigr)
    \end{equation}
    for all $A \in \Alg_\Lambdal$ and all $t \in I$, where $\rho_0^\Lambdak(t)$ denotes a suitable ground state of $H_0^\Lambdak(t)$.
    Indeed, assuming that $\rho_0^\Lambdak(t) \rightharpoonup\rho_0(t)$ for every $t \in I$,%
    \footnote{
        Note that the sequence $(\rho_0^\Lambdak(t))_{k \in \N}$ is compact for every fixed $t \in I$ (Banach-Alaoglu Theorem).
        Moreover, it is shown in Proposition~5.3.25 in~\cite{BR II 1996} that every limit point of a sequence of ground states associated to a converging sequence of derivations $\mathcal{L}^\Lambdak_{H_0(t)} \to \mathcal{L}_{H_0(t)}$ is a ground state of the limiting derivation.}
    this simply follows after taking the limit $k \to \infty$ on both sides of~\eqref{eq:finitegapinbulk} and realizing that, as $k \to \infty$, the set of admissible observables $A \in \Alg_{\Lambda_{l(k)}}$ exhausts $\Aloc$, which is dense in $\Alg_\Gamma$ by definition.
    The resulting inequality immediately yields the desired spectral gap for the GNS Hamiltonian (cf.~\cite[Proposition~5.3.19]{BR II 1996} and~\cite[Section~7]{nachtergaele2020quasi}).
\end{rmk}

In the following, we shall present adiabatic theorems for infinite systems (Theorem~\ref{thm:existenceofneass3}) as well as for extended but finite systems (Theorem~\ref{thm:existenceofneass4}) under Assumption~\nameref{ass:GAPbulk}.

\subsubsection{Infinite systems}
Analogously to Section~\ref{subsec:unifgap}, the basic assumptions on the Hamiltonian say that it is composed of exponentially localized interactions and/or a Lipschitz potential.
In addition, the Hamiltonian $H_0$ satisfies a technical regularity assumption in $t$, for which we recall that $I \subset \mathbb{R}$ denotes an open time interval.

\begin{assumption}{(INT\textsubscript{3})}{Assumptions on the interactions}
    \label{ass:INT3}
    \vspace{\topsep}
    \begin{itemize}
        \item[(i)] Let $\Psi_{H_0}, \Psi_{H_1} \in  \mathcal{B}_{I,a,\infty }^\circ$ be time-dependent  infinite volume interactions and  $v_\infty\in\mathcal{V}^\circ_I$ a time-dependent infinite volume Lipschitz potential.
        \item[(ii)] Assume that the map $I\to  \mathcal{B}_{a,\infty }^\circ$, $t\mapsto \Psi_{H_0(t)}$ is continuously differentiable.%
            \footnote{
                Note that this technical assumption does not follow from $\Psi_{H_0} \in \mathcal{B}_{I,a,\infty }^\circ$, as the spaces of smooth and bounded interactions are defined via term-wise and point-wise time derivatives (cf.~Section~\ref{subsec:interactions}).}
    \end{itemize}
\end{assumption}

We can now formulate the third generalized super-adiabatic theorem concerning infinite systems with a gap in the bulk~\cite{henheikteufel20203}.

\begin{theorem}
    \theoremtitlecite{Adiabatic theorem for infinite systems with a gap in the bulk}{\cite[Theorem~3.4]{henheikteufel20203}}
    \label{thm:existenceofneass3}
    Under Assumptions~\nameref{ass:GAPbulk} and~\nameref{ass:INT3}, there exists a near-identity automorphism $\beta^{\varepsilon, \eta}(t)= \E^{\I \varepsilon\mathcal{L}_{S^{\varepsilon, \eta}(t)}}$ on $\Alg_\Gamma$ with SLT generators $ S^{\varepsilon, \eta}$ for any $\varepsilon, \eta \in (0,1]$ and $t \in I$  such that the state
    \begin{equation*}
        \Pi^{\varepsilon, \eta}(t) := \rho_0(t)\circ \beta^{\varepsilon, \eta}(t)
    \end{equation*}
    is a {super-adiabatic NEASS} for $\rho_0(t)$ and the Heisenberg time-evolution on $\Alg_\Gamma$ generated by
    $\frac{1}{\eta}\Psi_{H^{\varepsilon}(\cdot)}$ with
    \begin{equation*}
        \Psi_{H^{\varepsilon}}:= \Psi_{H_0} + \epsi \,(V_{v_\infty} + \Psi_{H_1})\,.
    \end{equation*}
\end{theorem}

The key role of the spectral gap condition is that it allows to construct an inverse of the Liouvillian $\mathcal{L}_{H_0(t)}$, appearing in the construction of the NEASS, which maps SLT operators to SLT operators with slightly deteriorated locality properties.
Hence, the inverse of $\mathcal{L}_{H_0(t)}$ is called the \emph{quasi-local inverse of the Liouvillian}.%
\footnote{\label{fn:qliL}
    This particular phrase was used in~\cite{teufel2020non}.
    Others call it \emph{local inverse}~\cite{bachmann2018adiabatic,monaco2017adiabatic} or just \emph{inverse}~\cite{henheikteufel2020,henheikteufel20202,nachtergaele2019quasi}.
    In~\cite{henheikteufel20203}, it was called \emph{SLT inverse}.
    To avoid confusion, we want to reserve the SLT prefix for operators, i.e. SLT operator or SLT generator, but not for maps between SLT operators.
}
\def\footnoteInverseLiouvillian{\value{footnote}}
Assuming a gap only in the bulk, as done in \nameref{ass:GAPbulk}, means that
the action of the Liouvillian can only be \textit{inverted in the bulk} (see Section~\ref{sec:proof}).

\subsubsection{Extended but finite systems}
Contrary to the results in Section~\ref{subsec:unifgap}, the adiabatic theorem describing an infinite system with a gap in the bulk did \emph{not} require any notion of having a TDL in its formulation.
Instead, in order to derive a finite-volume analogue from Theorem~\ref{thm:existenceofneass3} (with qualitative additional error terms, see Theorem~\ref{thm:existenceofneass4} below), we need to introduce the stronger notion of \emph{having a rapid thermodynamic limit} for the exponentially localized interactions and the Lipschitz potential.
We refer to~\cite{henheikteufel20203} for a detailed discussion of this property.

\begin{defi}
    \theoremtitle{Rapid thermodynamic limit of interactions and potentials}
    \label{def:rtd}
    \vspace{\topsep}
    \begin{itemize}
        \item[(a)] An exponentially localized time-dependent interaction $\Phi  \in \mathcal{B}_{I,a, \infty }$ is said to \emph{have a rapid thermodynamic limit with exponent $\gamma \in (0,1)$} (have a \RTDLg{\gamma}) if
            there exists an infinite volume interaction $\Psi  \in \mathcal{B}_{I,a, \infty }^\circ$
            such that
            \begin{equation}\label{eq:rapidtdl}
                \begin{aligned}
                    \forall \, n \in \N, \, i \in \mathbb{N}_0 \ \ \exists \lambda,  C>0 &\ \ \forall M\in \N\ \ \forall \,   k \, \ge \, M + \lambda M^{\gamma} :
                    \\[1.5mm] &\sup_{t \in I}\left\Vert  \frac{\D^i}{\D t^i}\left( \Psi- \Phi^{ \Lambda_k}\right) (t) \, \right\Vert_{a, n,\Lambda_M}  \le C \, \E^{-a M^\gamma}\,,
                \end{aligned}
            \end{equation}
            and we write $\Phi\rtdlim \Psi$ in this case.

            A family of operators is said to \emph{have a \RTDL{}} if and only if the corresponding interaction does.

            For more general (non-exponentially localized) SLT operators, the definition is completely analogous.
        \item[(b)] 	A Lipschitz potential
            $v \in \mathcal{V}_I$
            is said to \emph{have a \RTDLg{\gamma}} if it is eventually independent of $k$, i.e.~if  there exists  an infinite volume Lipschitz potential $ v_\infty \in \mathcal{V}_I^\circ$ 		such that
            \begin{align*}
                \exists \lambda >0\ \ \forall M\in \N \ \ \forall \,  k \, \ge \, M + \lambda M^{\gamma}, t \in I: \ \  v_{ \infty}(t, \cdot) |_{\Lambda_M} = v^{ \Lambda_k}(t, \cdot) |_{\Lambda_M} \,.
            \end{align*}
            Again, we write $v\rtdlim v_\infty$ in this case.
    \end{itemize}
\end{defi}

In a nutshell, having a \RTDLg{\gamma} means that the interaction (or the Lipschitz potential) essentially agrees with a corresponding infinite volume object, up to terms located on a thin shell with relative size of order $k^{\gamma-1}$ right at the boundary of $\Lambdak$.
Note that, whenever $\Phi = \Psi$ for some infinite-volume interaction $\Psi$, or $v = v_\infty$ for some infinite volume Lipschitz potential $v_\infty$, both $\Phi$ and $v$ trivially have a \RTDLg{\gamma} (with any exponent $\gamma \in (0,1)$).

The following Theorem~\ref{thm:existenceofneass4} is deduced from Theorem~\ref{thm:existenceofneass3} by comparing the time evolution $\mathfrak{U}_{t,t_0}^{\epsi, \eta}$ and the near identity automorphism $\beta^{\epsi, \eta}$ in the definition of the NEASS on the infinite system $\Gamma$ with the same objects for large (but finite) systems $\Lambdak$.
Therefore, we will need to assume the existence of a rapid thermodynamic limit for the building blocks of the Hamiltonian~\eqref{eq:basicHamiltonian}.

\begin{assumption}{(INT\textsubscript{4})}{Assumptions on the interactions}
    \label{ass:INT4}\noindent
    The interactions $\Phi_{H_0}, \Phi_{H_1} \in  \mathcal{B}_{I, a,\infty } $  and the Lipschitz potential $v\in \mathcal{V}_I$ all have a \RTDL, i.e.\ $\Phi_{H_0} \rtdlim \Psi_{H_0}$, $\Phi_{H_1} \rtdlim \Psi_{H_1}$ and $v \rtdlim v_\infty$.
    The limiting objects $\Psi_{H_0}$, $ \Psi_{H_1}$ and $v_\infty$ satisfy Assumption~\nameref{ass:INT3}.
\end{assumption}

In Theorem~\ref{thm:existenceofneass4} we shall consider finite volume states $\rho_0^\Lambdak(t)$, which are close to the infinite volume ground state $\rho_0(t)$ away from the boundary in following sense.

\begin{assumption}{(S\textsubscript{bulk})}{Assumption on the convergence of states}\label{ass:Sbulk}\noindent
    The sequence $\big(\rho_0^{\Lambda_k}(t)\big)_{k\in \N}$   of  states  on $\Alg_{\Lambda_k}$ converges rapidly to $\rho_0(t)$ in the bulk:     there exist $C\in\R$, $m  \in \mathbb{N}$ and $h \in \mathcal{S}$ such that   for any finite $X \subset \Gamma$, $A \in \Alg_X$ and $\Lambda_k\supset X$
    \begin{equation*}
        \sup_{t \in I} \left\vert \rho_0(t)(A)  - \rho_0^{\Lambda_k}(t)(A) \right\vert
        \le
        C \, \Vert A \Vert \diam(X)^m \,  h\big(\operatorname{dist}(X,\Gamma \setminus \Lambda_k)\big)\,.
    \end{equation*}
\end{assumption}

While  the sequence $\rho_0^\Lambdak(t) \equiv \rho_0(t)|_{\Alg_{\Lambda_k}}$ of simple restrictions satisfies Assumption~\nameref{ass:Sbulk} trivially, the adiabatic theorem ensures the existence of a super-adiabatic NEASS constructed for \emph{any} such sequence.%
\footnote{
    This is why we wrote \emph{‘(close to) a ground state’} in Definition~\ref{def:NEASS}.}
Most interesting for physical application would be  a sequence of ground states
$\rho_0^{\Lambda_k}(t)$   of the finite volume Hamiltonians $H_0^{\Lambda_k}(t)$.
While the above assumption is expected to hold for any sequence of finite volume ground states for  Hamiltonians modeling Chern or topological insulators like in~\eqref{eq:examplehamiltonian2}, the only result we are aware of indeed \emph{proving} such a statement is again (see the discussion below Assumption~\nameref{ass:GAPbulk}) for weakly interacting spin systems~\cite{yarotsky2005uniqueness}.
In spirit, assuming \nameref{ass:Sbulk} is very similar to supposing that the system satisfies \emph{local topological quantum order} (LTQO)~\cite{MZ13, nachtergaele2021stability} or a strong \emph{local perturbations perturb locally} (LPPL) principle for perturbations acting at the boundary of the system~\cite{HTW2022, bachmann2021}.

We can now formulate the fourth and last generalized super-adiabatic theorem concerning finite systems with a gap in the bulk~\cite{henheikteufel20203}.

\begin{theorem}
    \theoremtitlecite{Adiabatic theorem for finite systems with gap in the bulk}{\cite[Theorem~4.1]{henheikteufel20203}}
    \label{thm:existenceofneass4}
    Under the Assumptions~\nameref{ass:GAPbulk}, \nameref{ass:INT4} and~\nameref{ass:Sbulk}, there exists a sequence of near-identity automorphisms $\beta^{\varepsilon, \eta, \Lambdak}(t)= \E^{\I \varepsilon\mathcal{L}^\Lambdak_{S^{\varepsilon, \eta}(t)}}$ with SLT generators $ S^{\varepsilon, \eta}$ for any $\varepsilon, \eta \in (0,1]$ and $t \in I$, such that the states
    \begin{equation*}
        \Pi^{\varepsilon, \eta, \Lambdak}(t) := \rho_0^\Lambdak(t)\circ \beta^{\varepsilon, \eta, \Lambdak}(t)
    \end{equation*}
    are super-adiabatic NEASSs for the Heisenberg time-evolution $\mathfrak{U}_{t,t_0}^{\varepsilon, \eta, \Lambda_k}$ on $\Alg_\Lambdak$ generated by $\tfrac{1}{\eta} \, H^{\varepsilon,\Lambda_k}(\cdot)$ with 
    \begin{equation*}
        \tfrac{1}{\eta} \, H^{\varepsilon,\Lambda_k}(t)
        := \tfrac{1}{\eta} \Bigl( H^{ \Lambda_k}_0(t) + \epsi \, \bigl(
        V_v^{\Lambda_k} (t) + H^{\Lambda_k}_1(t) \bigr)\Bigr)\,,
    \end{equation*}
    up to an error vanishing faster than any inverse polynomial in the distance to the boundary.
    That is, for any $n \in \mathbb{N}$  there exists a constant  $C_n$ {and for any compact $K\subset I$ and $m\in \N$ there exists a constant $\tilde{C}_{n,m,K}$} such that for all {$k\in \N$}, all $X \subset \Lambda_k$, all $A \in \Alg_X$ and all $t,t_0\in K$
    \begin{align} \nonumber
        &\left\vert \Pi^{\varepsilon, \eta, \Lambda_k}(t_0)(\mathfrak{U}_{t,t_0}^{\varepsilon, \eta, \Lambda_k}\Ab{A}) - \Pi^{\varepsilon, \eta, \Lambda_k}(t)(A) \right\vert  \\ \label{eq:adiabboundfinitebulkgap}
        & \qquad \qquad \qquad \le \ C_n \, \frac{\varepsilon^{n+1} + \eta^{n+1}}{\eta^{d+1}} \,  \left(1+\vert t- t_0\vert^{d+1}\right) \  \Vert A \Vert \, \vert X \vert^2 \\ \nonumber
        & \qquad \qquad \qquad \quad \quad + \ \tilde{C}_{ n,m,K} \,  \left(1+\eta \,  \operatorname{dist}(X,\Gamma \setminus \Lambdak )\right)^{-m}\Vert A \Vert \diam(X)^{2d}\,.
    \end{align}
\end{theorem}

The above theorem asserts that by assuming \nameref{ass:GAPbulk},
one obtains similar adiabatic bounds also for states of finite systems (without a spectral gap!)
which are close to the infinite volume ground state in the bulk as formulated in Assumption~\nameref{ass:Sbulk}.
Since adiabaticity potentially breaks at the boundaries of the finite systems, non-adiabatic effects arising close to the boundary may propagate into the bulk.
Therefore, an additional error term appears, but it decays faster than any polynomial in the size of the finite system for any  fixed $\eta$.
The actual form of the additional error term in the last line of~\eqref{eq:adiabboundfinitebulkgap} coming out of the proof in~\cite[Section~5]{henheikteufel20203} is slightly better but more complicated, which is why we refrain from stating it here.

The main points in the proof of Theorem~\ref{thm:existenceofneass4}, which we discuss in Section~\ref{sec:proof}, are to show that (i) the property of having a \RTDLg{\gamma} is preserved under all necessary operations for the construction of the NEASS (similarly as for Theorem~\ref{thm:existenceofneass2}) and (ii) having a \RTDLg{\gamma} for an interaction provides an explicit rate of convergence for the associated evolution family as in Proposition~\ref{tdlofcauchyinteractions}.

\section{Idea of the proofs}
\label{sec:proof}

The goal of the present section is to convey the main ideas relevant for proving the individual theorems from Section~\ref{sec:result}, where we already glimpsed the key steps required in their proofs.
For many technical details we refer the reader to the original works~\cite{teufel2020non,henheikteufel20202,henheikteufel20203}.

\subsection{Systems with a uniform gap}

The fundamental conceptual idea behind the proof for all four variants of the generalized super-adiabatic theorems is a perturbative scheme, which was called \emph{space-time adiabatic perturbation theory} in~\cite{PST2003,PST2003b}.
The basic structure of this computation is most easily presented for finite systems, where no further technical difficulties arise since all appearing operators are in fact matrices and thus bounded.
However, it is still necessary to show that all estimates are uniform in the size of the system $\Lambdak$.

\subsubsection{Extended but finite systems: Proof of Theorem~\ref{thm:existenceofneassfinite}}
\label{sec:proof-finite-systems-uniform-gap}

The form in which we presented Theorem~\ref{thm:existenceofneassfinite} differs slightly from the original result~\cite[Theorem~5.1]{teufel2020non}.
The original statement concerns a sequence \(\Pi_n^{\epsi,\eta,\Lambdak}(t) := \rho_0^{\Lambdak}(t)\circ \beta_n\eeL(t)\) of states on $\Lambdak$ (indexed by $n \in \N$), where
\begin{equation*}
    \beta_n\eeL(t)\Ab{A} := \E^{-\I \varepsilon \mathcal{L}^\Lambdak_{S_n\ee(t)}}\Ab{A}
    \quadtext{and}
    \varepsilon \mathcal{L}^\Lambdak_{S_n\ee(t)} := \sum_{j = 1}^{n} \varepsilon^{j} \mathcal{L}^\Lambdak_{A_j\ee(t)}\,.
\end{equation*}
From this, Theorem~\ref{thm:existenceofneassfinite} (and similarly all other three theorems) follows by a simple \emph{resummation} of the \(\varepsilon^{j}\smash{\mathcal{L}^\Lambdak_{A_j\ee(t)}}\), which will be discussed in Section~\ref{sec:proof-resummation} below.

The main idea of the proof is to choose each operator $A_j\eeL(t)$, $j =1, \dotsc , n$, in such a way that the \(j\)\textsuperscript{th}-order term in the perturbative scheme vanishes.
For the \(n\)-dependent result (i.e.~prior to resummation), we apply the fundamental theorem of calculus to get
\begin{equation}\label{eq:proof-fundamenta-theorem-of-calculus}
    \Pi_n\eeL(t_0)\bigl(\mathfrak{U}_{t,t_0}\eeL\Ab{A}\bigr) - \Pi_n\eeL(t)(A)
    = - \int_{t_0}^{t} \mathrm{d}s \ \frac{\mathrm{d}}{\mathrm{d}s} \,\rho_0^{\Lambdak}(s)  \Bigl(  \beta_n\eeL(s)\circ  \mathfrak{U}_{t,s}\eeL\Ab{A} \Bigr)
\end{equation}
and then aim to bound the integrand.
Calculating the derivative by using the chain rule and Duhamel's formula leaves us with
\begin{equation}\label{eq:proof-integrand-of-first-step}
    \frac{\mathrm{d}}{\mathrm{d}s} \,\rho_0^{\Lambdak}(s)  \Bigl(  \beta_n\eeL(s)\circ  \mathfrak{U}_{t,s}\eeL\Ab{A} \Bigr)
    =
    -\frac{\I}{\eta} \, \rho_0^{\Lambdak}(s)\Bigl(\Bigl[
        Q^{\epsi, \eta, \Lambdak}_n(s),
        \beta_n\eeL(s)\circ  \mathfrak{U}_{t,s}\eeL\Ab{A}
    \Bigr]\Bigr)\,,
\end{equation}
where $ Q^{\epsi, \eta, \Lambdak}_n(s)$ is a shorthand notation for
\begin{align}
    &\alignindent
    \begin{multlined}[t]
        \eta \,  \mathcal{I}_s^{\Lambdak}\bigl(\dot H_0^\Lambdak(s)\bigr)
        +\eta \int_0^1 \mathrm{d}\lambda \, \E^{-\I\lambda\varepsilon S_n\eeL(s)}\, \varepsilon \dot S_n\eeL(s)\, \E^{\I\lambda\varepsilon S_n\eeL(s)}
        \\+ \E^{-\I\varepsilon S_n\eeL(s)} \,\bigl( H_0^\Lambdak(s)+\varepsilon V^\Lambdak(s)\bigr) \,\E^{\I\varepsilon S_n\eeL(s)}
    \end{multlined}%
    \nonumber
    \\&=:
    H_0^\Lambdak(s) + \sum_{j=1}^n \varepsilon^j R_j\eeL(s) + \varepsilon^{n+1} R_{n+1}\eeL(s)\,,
    \label{eq:proof-integrand-expansion-in-epsilon}
\end{align}
and \(V^{\Lambda_k} = V_v^{\Lambda_k} + H_1^{\Lambda_k}\).
Here, \(\mathcal{I}_s^{\Lambdak}\bigl(\dot H_0^\Lambdak(s)\bigr)\) is the SLT generator of the parallel transport within the  vector-bundle $\Xi_{0,I}^\Lambdak$ over $I$ defined by $t \mapsto \rho_0^\Lambdak(t)$.%
\footnote{
    Since we assumed uniqueness of the ground state $\rho_0^\Lambdak(t)$, the vector-bundle $\Xi_{0,I}^\Lambdak$ is one-dimensional.
    If this were not the case, one had to include further terms generating the internal dynamic in $\Xi_{0,I}^\Lambdak$ (see~\cite{teufel2020non, henheikteufel20202}).}
This parallel transport is also known as the \emph{spectral flow}, which plays a fundamental role in proving automorphic equivalence of gapped ground state phases (see e.g.~\cite{bachmann2012automorphic,bachmann2018adiabatic}).
Moreover, the operator \(\mathcal{I}_s^{\Lambdak}\colon\Alg_{\Lambda_k} \to \Alg_{\Lambda_k}\) is called the \emph{quasi-local inverse of the Liouvillian}\textsuperscript{\ref{fn:qliL}} $\mathcal{L}_{H_0(s)}^\Lambdak$, since it satisfies~\cite{bachmann2018adiabatic,teufel2020non}
\begin{equation}\label{eq:proof-qliL-finite-lattices}
    \rho_0^\Lambdak(s)\Bigl(\bigl[\mathcal{L}_{H_0(s)}^\Lambdak \circ \mathcal{I}_s^{\Lambdak}\Ab{B_1} - \I\, B_1,B_2\bigr] \Bigr) = 0
    \quadtext{for all}
    B_1,B_2\in\Alg_\Lambdak\,, \quad s \in I\,,
\end{equation}
and also preserves good localization of its argument (in particular, it maps SLT operators to SLT operators).
This combined property of $\mathcal{I}_s^\Lambdak$ heavily relies on the ground state $\rho_0^\Lambdak(s)$ being gapped~\cite{bachmann2018adiabatic, teufel2020non, nachtergaele2019quasi} and will be of fundamental importance in the following.

In the last line of~\eqref{eq:proof-integrand-expansion-in-epsilon}, we expanded in powers of \(\varepsilon\) and \(\eta\) in the sense that \(R_j\eeL(s)\), for \(j \le n\), are polynomials in \(\eta/\varepsilon\) of order (at most) \(j\) with \(\varepsilon\)- and \(\eta\)-independent SLT operators as coefficients.
A more detailed step-by-step calculation can be found in the proof of Proposition~5.1 in~\cite{teufel2020non}.
Let us here only report the general structure
\begin{equation} \label{eq:proof Rj structure}
    R_j\eeL(s)
    = -\I\, \mathcal{L}_{H_0(s)}^\Lambdak\bigl(A_j\eeL(s)\bigr)
    + \tilde R_j\eeL(s),
\end{equation}
where the first remainder term is given by
\begin{equation*}
    \tilde R_1\eeL(s)
    = \tfrac{\eta}{\varepsilon} \,\mathcal{I}_s^{\Lambdak}\bigl(\dot H_0^\Lambdak(s)\bigr)-V^\Lambdak(s)
\end{equation*}
and all other \(\tilde R_j\eeL(s)\) are composed of iterated commutators of the operators \(A_i\eeL(s)\) and \(\dot A_i\eeL(s)\), for \(i<j\le n\), with \(H_0^\Lambdak(s)\) and \(V^\Lambdak(s)\).
In contrast to general onsite potentials, the commutator of a Lipschitz potential with an SLT operator is an SLT operator itself~\cite[Lemma~2.1]{teufel2020non}.
For the commutator of SLT operators, this is easy to see.

We now consider individual terms from~\eqref{eq:proof-integrand-expansion-in-epsilon} when plugged into~\eqref{eq:proof-integrand-of-first-step}.
The zero-order term vanishes, because \(\rho_0^\Lambdak(s)\) is the ground state of \(H_0^\Lambdak(s)\).
By application of~\eqref{eq:proof-qliL-finite-lattices} we can iteratively choose
\begin{equation} \label{eq:proof Aj structure}
    A_j\eeL(s) = -\mathcal{I}_s^\Lambdak\bigl(\tilde R_j\eeL(s)\bigr)
\end{equation}
such that~\eqref{eq:proof-integrand-of-first-step} vanishes up to
\begin{equation} \label{eq:proof-remainder}
    -\I\,\frac{\varepsilon^{n+1}}{\eta}\,\rho_0^{\Lambdak}(s)\Bigl(\Bigl[
        R_{n+1}\eeL(s),
        \beta_n\eeL(s)\circ  \mathfrak{U}_{t,s}\eeL\Ab{A}
    \Bigr]\Bigr)\,.
\end{equation}

Moreover, all the operations involved in calculating the \(A_j\eeL(s)\), i.e.\ taking commutators and applying the quasi-local inverse of the Liouvillian preserve the locality properties of the operators as shown in the appendices of~\cite{teufel2020non,monaco2017adiabatic}, which are heavily based on~\cite{bachmann2018adiabatic}.
Hence, also all \(A_j\eeL\) are SLT operators.

It turns out that also \(R_{n+1}\eeL\) is a polynomial in \(\eta/\varepsilon\) of order at most \(n+1\) and its coefficients, as we just explained, are SLT operators~\cite[Proof of Proposition~6.1]{teufel2020non}.
Thus, the absolute value of~\eqref{eq:proof-fundamenta-theorem-of-calculus} is bounded by
\begin{align}
    &\alignindent\frac{\varepsilon^{n+1}}{\eta}\, \biggl\lvert \int_{t_0}^{t} \mathrm{d} s
    \,\rho_0^{\Lambdak}(s)\Bigl(\Bigl[
        R_{n+1}\eeL(s),
        \beta_n\eeL(s)\circ  \mathfrak{U}_{t,s}\eeL\Ab{A}
        \Bigr]\Bigr) \biggr\rvert \nonumber
    \\[1mm]&\leq
    \frac{\varepsilon^{n+1}}{\eta}
    \, \lvert t-t_0 \rvert
    \sup_{s\in [t_0,t]} \Bigl\lVert \Bigl[ \bigl(\beta_n\eeL\bigr)^{-1}(s) \bigl(R_{n+1}\eeL(s)\bigr), \mathfrak{U}_{t,s}\eeL\Ab{A} \Bigr] \Bigr\rVert \nonumber
    \\[1mm]&\leq
    C_n \,\frac{\varepsilon^{n+1}}{\eta}
    \, \lvert t-t_0 \rvert
    \, \biggl(1+\Bigl(\frac{\mathstrut\eta}{\varepsilon}\Bigr)^{n+1}\biggr)
    \biggl(1+\Bigl(\frac{\lvert t-t_0 \rvert}{\eta}\Bigr)^{d}\biggr)
    \, \lVert A \rVert
    \, \lvert X \rvert^2 \label{eq:remainder int est}
    \\[1mm]&\leq
    C_n \, \frac{\varepsilon^{n+1}+\eta^{n+1}}{\eta^{d+1}}
    \, \lvert t-t_0 \rvert
    \, \bigl(1+\lvert t-t_0\rvert^d\bigr)
    \, \lVert A \rVert 
    \, \lvert X \rvert^2\,, \nonumber
\end{align}
where we essentially used a generalized Lieb-Robinson bound~\cite[Lemma~B.5]{teufel2020non} to estimate the commutator.
Note that the \((1+\bigl(\lvert t-t_0 \rvert / \eta\bigr){}^d)\)-factor comes from the Lieb-Robinson bound and the adiabatic \(1/\eta\)-scaling of the time evolution \(\mathfrak{U}_{t,s}\).
The \((1+(\eta/\varepsilon)^{n+1})\)-factor comes from bounding the interaction norm of \(R_{n+1}\eeL(s)\) by separating the polynomial dependence on \(\eta/\varepsilon\) such that \(C_n\) is independent of \(\Lambdak\), \(\varepsilon\) and \(\eta\).
We have thus shown that the NEASS almost intertwines the time evolution, i.e.\ item~\ref{def:NEASS-general-bound} of Definition~\ref{def:NEASS}.

We are left with discussing the remaining three characterizing properties of the NEASS given in Definition~\ref{def:NEASS}:
By construction, all \(A_j\eeL(t)\) depend only on \(H_0^\Lambdak(t)\) and \(V^\Lambdak(t)\) and their \(j\)\textsuperscript{th} derivatives at time \(t\).
This shows that the NEASS is local in time, i.e.\ item~\ref{def:NEASS-local-in-time}.
Moreover, if all time derivatives of \(H_0\) and \(V\) vanish for some \(t\in I\), then all non-constant (i.e.\ in front of some positive power of \(\eta/\varepsilon\)) coefficients in \(R_j\eeL\) vanish and \(\Pi_n\eeL(t)=\Pi_n^{\varepsilon,0,\Lambdak}(t)\).
This shows that the NEASS is stationary whenever the Hamiltonian is stationary, i.e.\ item~\ref{def:NEASS-stationary}.
If, for some \(t\in I\), \(\dot H_0^\Lambdak(t)\) and \(V^\Lambdak(t)\) vanish, then \(\tilde R_1\eeL\) and thus \(A_1\eeL\) vanish.
If additionally all derivatives of \(H_0^\Lambdak\) and \(V^\Lambdak\) at \(t\) vanish, also \(\tilde{R}_j^\Lambdak(t)\) and thus \(A_j\eeL(t)\) vanish.
Hence, \(\beta_n\eeL(t)=\unit_\Lambdak\) and the NEASS equals the ground state, i.e.\ item~\ref{def:NEASS-equals-ground-state} holds.

The above listed general arguments immediately translate to the other three theorems.

\subsubsection{Infinite systems: Proof of Theorem~\ref{thm:existenceofneass2}}
\label{sec:proof-infinite-systems-uniform-gap}

Without any further assumptions, the sequence Hamiltonian $H^{\varepsilon,\Lambdak}$ and its constituents $H_0^\Lambdak$ and $V^\Lambdak$ could have nothing in common for different lattice sizes $k$ (they might even describe different physical systems), so taking the limit $\Lambdak \nearrow \Gamma$ might not be well-defined.
In order to avoid this somewhat meaningless situation, we assumed that the building blocks of the Hamiltonian \emph{have a TDL} (see Definition~\ref{def:td} and Assumption~\nameref{ass:INT2}) and also the sequence of ground states $\big(\rho_0^\Lambdak(t)\big)_{k \in \N}$ converges (Assumption~\nameref{ass:Sunif}).
Since the property of having a TDL guarantees the existence of the thermodynamic limit for the corresponding evolution operators (see Proposition~\ref{tdlofcauchyinteractions} and~\cite{nachtergaele2019quasi}), it remains to show that the operator sequences \(\big(A_j\eeL(t)\big)_{k \in \N}\), $j = 1, \dotsc , n$, constructed in Section~\ref{sec:proof-finite-systems-uniform-gap} have a TDL\@.
More precisely, one needs to show that taking time-derivatives, sums of commutators with the building blocks of $H^\epsi$ (and $\dot{H}_0$), and the inverse of the Liouvillian (see~\eqref{eq:proof Rj structure}) leaves the property of having a TDL for SLT operators invariant, which is in fact the main point of the proof in~\cite{henheikteufel20202}.
It is then straightforward to show that compositions of states and automorphisms, all having a thermodynamic limit, converge as $\Lambdak \nearrow \Gamma$.
Since the constant $C_n$ from~\eqref{eq:remainder int est} is uniformly bounded in $k$, the (sketch of a) proof of Theorem~\ref{thm:existenceofneass2} is complete.

\subsection{Systems with a gap in the bulk}

For systems having a spectral gap only in the bulk (i.e.~for the GNS Hamiltonian), the characteristic~\eqref{eq:proof-qliL-finite-lattices} of $\mathcal{I}_s^\Lambdak$, that it essentially inverts the Liouvillian $\mathcal{L}_{H_0(s)}^\Lambdak$ (and still maps SLT operators to SLT operators), is now only fulfilled for certain $B_1$ and $B_2$ in a dense domain \(\mathcal{D}\subset \Alg_\Gamma\) after taking the limit $\Lambdak \nearrow \Gamma$ (see~\cite[Proposition~3.3]{henheikteufel20203}).
Presuming that the limit actually exists, this point is the main challenge in proving an adiabatic theorem under the less restrictive gap Assumption~\nameref{ass:GAPbulk}.

\subsubsection{Infinite systems: Proof of Theorem~\ref{thm:existenceofneass3}}
\label{sec:proof-infinite-systems-bulk-gap}

As just explained, the main difficulty in proving Theorem~\ref{thm:existenceofneass3} is that~\eqref{eq:proof-qliL-finite-lattices} only holds if \(H_0^\Lambdak\) is gapped.
On top of that, we cannot handle the limit \(\Lambdak\nearrow\Gamma\) of the \(\tilde R_j\eeL\) directly nor could they be used in the infinite volume version of~\eqref{eq:proof-qliL-finite-lattices} because it only holds for \(B_1,B_2\in \mathcal{D}\subset \Alg_\Gamma\).
However, the rest of the construction from Section~\ref{sec:proof-finite-systems-uniform-gap} is still valid, but the lower order terms in~\eqref{eq:proof-fundamenta-theorem-of-calculus} have a non-vanishing contribution in finite domains.
We thus repeat this construction but take coefficients \(A_j\eeLL(t)\), which are built up from \(H_0^\Lambdak(t)\) but restricting the perturbations $\dot{H}_0(t)$ and $V(t)$ to \(\Lambdal\) with \(l<k\).
In this way, one can take the limit \(\Lambdak\nearrow\Gamma\) in~\eqref{eq:proof-qliL-finite-lattices} with \(B_1 = \lim_{k\to\infty} \tilde R_j\eeLL \in \Alg_\Gamma\) (see~\eqref{eq:proof first} and~\eqref{eq:proof higher} and the comment thereafter for technical obstructions in taking the limit).
Using this notational convention, we introduce the states
\begin{equation*}
    \Pi_n\eeLL(t) = \rho_0(t) \circ \beta\eeLL(t)\,,
\end{equation*}
where $\rho_0(t)$ is the infinite volume ground state, and compare them to the actual objects in infinite volume while estimating
\begin{align}
    \Bigl\lvert \Pi_n\ee(t_0)\bigl(\mathfrak{U}_{t,t_0}\ee\Ab{A} \bigr)-\Pi_n\ee(t)(A)\Bigr\rvert
    \le{}&
    \Bigl\lvert \Pi_n\ee(t_0)\bigl(\mathfrak{U}_{t,t_0}\ee\Ab{A}\bigr) -\Pi_n\eeLL(t_0)\bigl(\mathfrak{U}_{t,t_0}\eeLL\Ab{A}\bigr)  \Bigr\rvert \nonumber 
    \\&+
    \Bigl\lvert \Pi_n\eeLL( t_0)\bigl(\mathfrak{U}_{t,t_0}\eeLL\Ab{A}\bigr) - \Pi_n\eeLL( t)(A)\Bigr\rvert \label{eq:proof-bulk-split-finite-volume}
    \\&+
    \Bigl\lvert \Pi_n\eeLL( t)(A) -\Pi_n^{}( t)(A)\Bigr\rvert  \nonumber 
\end{align}
by means of the triangle inequality.
Since all the interactions (and the Lipschitz potential) have a TDL, one can prove~\cite[Section~5.1(b)]{henheikteufel20203} that the first and last summand in~\eqref{eq:proof-bulk-split-finite-volume} can be made arbitrarily small for \(k, l \in \N\) large enough, and we can thus focus on the second summand.
However, since~\eqref{eq:proof-qliL-finite-lattices} only holds in the limit $\Lambdak \nearrow \Gamma$ and also \(\rho_0(t)\) is not necessarily a ground state of \(H_0^\Lambdak(t)\), the lower order terms in the analogues of~\eqref{eq:proof-integrand-of-first-step} and~\eqref{eq:proof-integrand-expansion-in-epsilon} do not vanish for finite~\(k\) and~\(l\).
Instead, only
\begin{align}
    \lim\limits_{k \to \infty }\rho_0(s)\Bigl(\bigl[ H_0^\Lambdak(s), \beta_n\eeLL(s)  \circ \mathfrak{U}_{t,s}\eeLL\Ab{A}\bigr]\Bigr) = 0 \label{eq:proof first}
    \shortintertext{and}
    \lim\limits_{k \to \infty }\rho_0(s)\Bigl(\bigl[R_j\eeLL(s), \beta_n\eeLL(s)  \circ \mathfrak{U}_{t,s}\eeLL\Ab{A}\bigr]\Bigr) = 0 \label{eq:proof higher}
\end{align}
for all \(l\in\N\) and uniformly for \(s\) and \(t\) in compacts.
These statements require a careful analysis of deteriorating localization properties along the expansion as well as convergence estimates in norms measuring the quality of localization (cf.~the norm $\Vert \cdot \Vert_f$ introduced in Assumption~\nameref{ass:GAPbulk}~(iii)), such that the limits really converge to the infinite volume version of~\eqref{eq:proof-qliL-finite-lattices} with \(B_1\) and \(B_2\) in a dense domain \(\mathcal{D}\subset \Alg_\Gamma\).
For further details, we refer to Proposition~3.2 and the statements in Appendix~B of~\cite{henheikteufel20203}, which are adaptions of technical estimates that were originally established for the proof of automorphic equivalence with a gap only in the bulk~\cite{MO2020}.
Now, combining~\eqref{eq:proof first} and~\eqref{eq:proof higher} with the estimates on the first and third summand in~\eqref{eq:proof-bulk-split-finite-volume}, we conclude that all the lower order terms vanish in the limit \(k \to \infty\) followed by $l \to \infty$, which finishes our sketch of the proof of Theorem~\ref{thm:existenceofneass3}.

\subsubsection{Extended but finite systems: Proof of Theorem~\ref{thm:existenceofneass4}}
\label{sec:proof-finite-systems-bulk-gap}

Let us briefly explain the strategy to prove Theorem~\ref{thm:existenceofneass4}.
In order to show~\eqref{eq:adiabboundfinitebulkgap}, we first estimate
\begin{align}
    \Bigl\lvert \Pi_n\eeL(t_0)\bigl(\mathfrak{U}_{t,t_0}\eeL\Ab{A}\bigr) - \Pi_n\eeL(t)(A) \Bigr\rvert
    \le{}&
    \Bigl\lvert\Pi_n\eeL(t_0)\bigl(\mathfrak{U}_{t,t_0}\eeL\Ab{A}\bigr) - \Pi_n\ee(t_0)\bigl(\mathfrak{U}_{t,t_0}\ee\Ab{A}\bigr) \Bigr\rvert \nonumber
    \\&+
    \Bigl\lvert \Pi_n\ee(t_0)\bigl(\mathfrak{U}_{t,t_0}\ee\Ab{A}\bigr) - \Pi_n\ee(t)(A) \Bigr\rvert \label{eq:proof finite bulk gap}
    \\&+
    \Bigl\lvert\Pi_n\eeL(t)(A) - \Pi_n\ee(t)(A) \Bigr\rvert \nonumber
\end{align}
and treat the three summands separately.
The second summand corresponds to the infinite system and can be estimated by means of Theorem~\ref{thm:existenceofneass3}, such that it accounts for the first contribution on the RHS of~\eqref{eq:adiabboundfinitebulkgap}.
We are left with bounding the remaining two summands in~\eqref{eq:proof finite bulk gap}.
These contribute the additional error term on the RHS of~\eqref{eq:adiabboundfinitebulkgap}.
To estimate them, we need explicit control on the speed of convergence (it must be faster than any inverse polynomial) for the states (see Assumption~\nameref{ass:Sbulk}) and automorphisms \(\beta_n\eeL\) and \(\mathfrak{U}_{t,t_0}\eeL\).
For the time evolution \(\mathfrak{U}_{t,t_0}\eeL\), the rapid convergence to \(\mathfrak{U}_{t,t_0}\ee\) is ensured by supposing that the building blocks of $H^\epsi$ have a RTDL (see Definition~\ref{def:rtd} and Assumption~\nameref{ass:INT4}).
This was  carried out in~\cite[Appendix~B]{henheikteufel20203}, building on estimates from~\cite[Section~3]{nachtergaele2019quasi}.
We remark that the adiabatic $1/\eta$-scaling of the time evolution is responsible for the factor $\eta$ appearing in the additional error term in~\eqref{eq:adiabboundfinitebulkgap}.
In order to show that also \(\beta_n\eeL \to \beta_n\ee\) sufficiently fast, we need to show that all \(A_j\ee\) have a RTDL, i.e.~the operations involved in constructing the generator of $\beta_n\ee$ leave the property of having a RTDL (essentially) invariant (see~\cite[Appendix~C]{henheikteufel20203}).
This finishes the sketch of the proof of Theorem~\ref{thm:existenceofneass4} and we refer to~\cite[Section~5.2]{henheikteufel20203} for further details.

\subsection{Resummation of the NEASS}
\label{sec:proof-resummation}

As mentioned in the beginning of Section~\ref{sec:proof-finite-systems-uniform-gap}, the statements formulated in Section~\ref{sec:result} require a resummation, which we explain in the following.
First, note that the generator $\varepsilon S_n\ee$ of \(\beta_n\ee\) constructed above can be rewritten as \(\varepsilon S_n\ee = \sum_{j=1}^n \sum_{i=0}^j \varepsilon^i \eta^{j-1} A_{j,i}\), where the coefficients \(A_{j,i}\) are time-dependent SLT operators and independent of \(\varepsilon\) and \(\eta\).
Now, it is easy to show (see~\cite[Lemma~E.1]{henheikteufel20202}) that there exists a sequence \(\delta_j \to 0\) and constants \(C_n\) such that the resummed generator
\begin{equation} \label{eq:resumm_S}
    \varepsilon S\ee = \sum_{j=1}^{\infty} \chi_{[0,1]}(\varepsilon / \delta_j) \,\chi_{[0,1]}(\eta / \delta_j) \sum_{i=0}^{j}\varepsilon^{i} \eta^{j-i} A_{j,i}
\end{equation}
satisfies
\begin{equation*}
    \bigl\lVert \varepsilon S\ee - \varepsilon S_n\ee \bigr\rVert_{\mathrm{SLT}} \leq C_n \, (\varepsilon^n + \eta^n)\,,
\end{equation*}
where $\Vert \cdot \Vert_{\mathrm{SLT}}$ denotes an interaction norm similar to~\eqref{normdefinition}.
Resummations of this type are standard, e.g., in microlocal analysis~\cite{Martinez} and the above estimate immediately leads to the bounds (cf.~\cite[Lemmata~E.3,~E.4]{henheikteufel20202})
\begin{align}
    \sup_{t \in I}\Bigl\lvert \Pi\eeL(t)(A) - \Pi_n\eeL(t)(A) \Bigr\rvert
    &\le
    C'_n \, (\varepsilon^{n}+\eta^n) \,\lVert A \rVert \,\lvert X \rvert^2 \label{eq:resumm 1}
    \quad\text{and}\\
    \Bigl\lvert \Pi\eeL(t_0)\bigl( \mathfrak{U}_{t,t_0}\eeL\Ab{A}\bigr) - \Pi_n\eeL(t_0)\bigl( \mathfrak{U}_{t,t_0}\eeL\Ab{A}\bigr)\Bigr\rvert
    &\le
    C''_n \, \frac{\varepsilon^{n} + \eta^n}{\eta^{d+1}}\, (1+\lvert t - t_0\rvert)^{d+1}\, \lVert A \rVert \, \lvert X \rvert^{2}\,,
    \label{eq:resumm 2}
\end{align}
uniformly in the size of the system $\Lambdak$. In the context of Theorem~\ref{thm:existenceofneass2} and Theorem~\ref{thm:existenceofneass3}, corresponding estimates hold in infinite volume, i.e.~without the subscript~$\Lambdak$. 

Next, since the sum in~\eqref{eq:resumm_S} is finite for every fixed $\epsi > 0$, also the resummed generator \(S\eeL\) has a TDL as soon as \(S_n\eeL\) has a TDL\@.
Therefore, the states \(\Pi\eeL\) constructed using the \(S\eeL\) instead of the \(S_n\eeL\) have a well-defined thermodynamic limit \(\Pi\ee\) (see~\cite[Lemma~E.2]{henheikteufel20202}) and since the bounds~\eqref{eq:resumm 1} and~\eqref{eq:resumm 2} are independent of \(\Lambdak\), they also hold for the respective objects in the thermodynamic limit.
Hence, the results formulated in Section~\ref{sec:result} can be concluded by combining the $n$-dependent statements discussed earlier in this section with the bounds~\eqref{eq:resumm 1} and~\eqref{eq:resumm 2} (or their infinite volume correspondents). 

\statement{Acknowledgments}
It is a pleasure to thank Stefan Teufel for numerous interesting discussions, fruitful collaboration and many helpful comments on an earlier version of the manuscript.
J.~H.~acknowledges partial financial support by the ERC Advanced Grant ‘RMTBeyond’ No.~101020331.

\statement{Conflict of interest}
The authors have no conflicts to disclose.

\statement{Author contributions}
\emph{Joscha~Henheik:} writing -- original draft (equal); writing -- review and editing (equal).
\emph{Tom~Wessel:} writing -- original draft (equal); writing -- review and editing (equal).

\statement{Data availability}
Data sharing is not applicable to this article as no new data were created or analyzed in this study.

\end{document}